\documentclass[symmetry,article,accept,pdftex,moreauthors]{Definitions/mdpi} 
\usepackage[labelformat=simple]{subfig}

\firstpage{1} 
\pubvolume{1}
\issuenum{1}
\articlenumber{0}
\pubyear{2025}
\copyrightyear{2025}
\externaleditor{Firstname Lastname} % More than 1 editor, please add `` and '' before the last editor name
\datereceived{4 October 2024 } 
\daterevised{29 January 2025 } % Comment out if no revised date
\dateaccepted{30 January 2025 } 
\datepublished{ } 
\hreflink{https://doi.org/} % If needed use \linebreak
\Title{Fundamental Oscillation Modes in Neutron Stars with Hyperons and Delta Baryons}
\TitleCitation{Fundamental Oscillation Modes in Neutron Stars with Hyperons and Delta Baryons}
\Author{O. P. Jyothilakshmi
 $^{1}$, P. E. Sravan Krishnan $^{2}$, V. Sreekanth $^{1,}$*,
Harsh Chandrakar $^{3}$ 
and Tarun Kumar Jha $^{3}$}

\AuthorNames{O. P. Jyothilakshmi, P. E. Sravan Krishnan, V. Sreekanth, Harsh Chandrakar and T. K. Jha}

\isAPAStyle{%
       \AuthorCitation{Lastname, F., Lastname, F., \& Lastname, F.}
         }{%
        \isChicagoStyle{%
        \AuthorCitation{Jyothilakshmi, O.P.; Krishnan, P.E.S.; Sreekanth, V.; Chandrakar, H.; Jha, T.K.}
        }{
        \AuthorCitation{Jyothilakshmi, O.P.; Krishnan, P.E.S.; Sreekanth, V.; Chandrakar, H.; Jha, T.K.}
        }
}

\address{%
$^{1}$ \quad Department of Physics, Amrita School of Physical Sciences, Amrita Vishwa Vidyapeetham, Coimbatore 641112, India; op\_jyothilakshmi@cb.students.amrita.edu 
\\
$^{2}$ \quad Argelander-Institut f\"ur Astronomie, Universit\"at Bonn, Auf dem H\"ugel 71, 53121 Bonn, Germany; s27skris@uni-bonn.de\\
$^{3}$ \quad Department of Physics, BITS PILANI K K Birla Goa Campus, Goa 403726, India; p20220023@goa.bits-pilani.ac.in~(H.C.); tkjha@goa.bits-pilani.ac.in~(T.K.J.)}

\corres{Correspondence: v\_sreekanth@cb.amrita.edu}

\abstract{For a new parameterization of the modified effective chiral model, developed primarily to regulate the density content of the symmetry energy and its higher order terms, equations of state (EoSs) for hyperon-rich matter ($H$) and delta baryon matter ($\Delta$) were obtained. The models were used to investigate the emission of gravitational waves (GWs) through $f$-mode oscillations in the corresponding neutron stars. We obtained the stellar structure, $f$-mode frequency and tidal deformability $\Lambda$ for our models. We report that the $\Delta$ EoS is stiffer compared to the $H$ EoS. We also analyzed the velocity of sound in these media. The corresponding mass--radius relationships were obtained and compared with various observations. We studied the dependence of $f$-mode frequencies on the stellar mass, redshift and tidal deformability. We employed the well known Cowling approximation to obtain the $f$-mode frequencies for $l=2,\,3$ and $4$ modes of oscillation. We found that the $f$-mode frequencies of the $H$ and $\Delta$ EoSs were almost the same in the lower mass region, while we observed a substantial difference between them in the high-mass region. 
We also obtained an empirical relation for the EoSs considered. The various attributes obtained for our models showed close agreement with various observational constraints from pulsars and GW events.}

\keyword{Neutron stars, Oscillations, Equations of state}%Use showkeys class option if keyword
\begin{document}

\section{Introduction}\label{Introduction} 
The detection of gravitational waves (GWs) from binary neutron star mergers~\cite{Abbott2017,LIGOScientific:2018cki,Abbott2020} has opened up a new way to constrain the equation of state (EoS) of dense matter found inside neutron stars. This is in addition to the basic constraint of meeting the maximal mass criteria set by the most massive of the observed pulsars, examples of which are PSR J0348-0432~\cite{Antoniadis2013}, PSR J1614-2230~\cite{Fonseca2016} and PSR J0740+6620~\cite{NANOGrav:2019jur}. Neutron stars (NSs) could experience a significant loss in angular momentum through the emission of GWs, which could potentially be detected by the upcoming Einstein Telescope or even by the Advanced LIGO/VIRGO.

\par
External perturbations can send NSs into oscillation in various quasi-normal modes, and these modes emit gravitational waves. 
Well-known NS oscillation modes include the fundamental ($f$) modes, pressure ($p$) modes, rotational ($r$) modes and gravitational ($g$) modes. Gravitational radiation can cause both the $f$- and $r$-modes to become unstable through the Chandrasekhar--Friedman--Schutz mechanism~\cite{CFS,Friedman78}, and this instability can be damped by dissipative effects such as the shear and bulk viscosities~\cite{Nayyar:2005th, Passamonti:2011eh,Doneva:2013zqa,Gittins:2022rxs,Jyothilakshmi:2022hys,Laskos-Patkos:2023cts}. 
However, it was found that unlike with $r$-modes, the~damping of $f$-modes leaves only a small region of instability~\cite{Doneva:2013zqa} in the plot between the angular velocity $\Omega$ and core temperature $T$. 
An unstable $f$-mode could have enough strength to be detected by gravitational wave observatories~\cite{Passamonti2013}. If~detected, these gravitational waves could provide a better understanding of the nature of the matter within~NSs. 
\par

The study of these oscillation modes requires solving the perturbed fluid equations in general relativity. The~first integrated numerical solution for $f$-modes was obtained by the authors of Ref.~\cite{Lindblom:1983ps} for various EoSs. The~relativistic Cowling approximation~\cite{Cowling:1941nqk}, obtained by neglecting the metric perturbations while considering the fluid oscillations, is widely used to study $f$-mode oscillations in compact stars. This approximation was found to differ from the general relativistic treatment by less than $20\%$~\cite{Sotani:2020mwc}.  It was used to study non-radial oscillations in compact stars in Refs.~\cite{PhysRevD.83.024014,Flores:2013yqa,Ranea-Sandoval:2018bgu}. More recently, the~Cowling approximation was used to estimate the $f$-mode frequencies for various other stellar configurations, such as bosonic dark matter~\cite{VasquezFlores:2019eht}, hyperon~\cite{Pradhan:2020amo}, hybrid~\cite{Kumar:2021hzo}, dark matter-admixed hyperon~\cite{Das:2021dru} and dark energy~\cite{Jyothilakshmi:2024zqn} stars. An~important constraint on the EoS is given by the analysis of the gravitational wave signal emitted during a binary inspiral, which is characterized by the tidal deformability ($\Lambda$). 
It measures the quadrupole deformation of the star due to the tidal field of a companion star. This is over and above the constraints already provided by the electromagnetic observations of neutron stars, which include their masses, radii, gravitational redshift and spin among others (see Ref.~\cite{Lattimer:2006xb}). The~GW170817 measurements provided limits on the radius and tidal deformability of a canonical NS ($1.4M_\odot$) in the range of $R_{1.4}=$ 12--13.5 km and $\Lambda_{1.4}=190_{-120}^{+390}$, respectively~\cite{LIGOScientific:2018cki}. %LIGOScientific:2018hze}. 
Additionally, a~range for the $f$-mode frequency was obtained as 1.67--2.18 kHz~\cite{Wen:2019ouw}. Future measurements from binary NS mergers may impose more strict limits on $f$-mode~frequencies. 

\par
The present model was treated very similarly to the well-known relativistic mean field approach to account for many baryon systems. Introduced by Gell-Mann and Levy~\cite{Gell-Mann:1960mvl}, it was subsequently applied to nuclear matter studies~\cite{Lee:1974ma}. After~a series of modifications~\cite{Lee:1974ma, Lee:1974uu, Furnstahl:1993wx, Serot:2002ei, Heide:1993yz, Mishustin:1993ub, Papazoglou:1998vr}, the~model was used to describe finite nuclear properties~\cite{Schramm:2002xi, Tsubakihara:2006se, Tsubakihara:2009zb}. Further, the~inclusion of the dynamically generated mass of the vector meson in the model~\cite{Boguta:1983uz, Sahu:1993db} resulted in an unrealistically high nuclear incompressibility ($K$) value, which was reduced to an acceptable range by introducing higher order terms of the scalar meson field and~then applied to nuclear matter studies~\cite{Jha2008}. 
Lately, mesonic cross-couplings have been incorporated~\cite{Malik2017} to regulate the density dependence of the symmetry energy parameters, and this was applied to study magnetized neutron stars~\cite{Patra2020}. It is to be noted that higher order interactions in the chiral fields are desirable as they are known to mimic three-body forces, which may play a vital role in dense matter~studies.

\par

With the ever-growing observations of massive stars, the~role of exotic matter such as quarks, hyperons, etc., in~compact stars has become interesting and important, particularly since very little is known about nucleon--hyperon, hyperon--hyperon or delta's interaction in matter. The observation of high-mass pulsars severely constrains the EoS and therefore the particle interactions, particularly when one considers exotics like hyperons or quarks in their inner shells. Their presence is known to have a softening effect on the EoS on one hand, while on the other hand we require a stiffer EoS in order to achieve a massive neutron star configuration. There are several prescriptions available in the literature to obtain high-mass stars; for~example, one may consider the effect of repulsive hyperon--hyperon interactions~\cite{Bednarek2012,Oertel2015} or repulsive hyperonic three-body forces~\cite{Vidana2000,Yamamoto2014,Lonardoni2013} or even a phase transition to deconfined quark matter~\cite{wei2019,Klahn2013}. 

In this work, we obtained two different models of neutron stars within the effective chiral model~\cite{Malik2017}, namely, one which includes the octet of baryons ($n, p, \Lambda^{0}, \Sigma^{-,0,+}, \Xi^{-,0}$) and another with delta baryons ($n, p,$
 $\Delta^{++,+,0,-}$). 
There have been studies on the impact of the $\Delta$ baryons of NSs within the density-dependent relativistic mean field formalism on various stellar oscillations like radial~\cite{Rather:2023dom,Rather:2024hmo} and non-radial~\cite{Kalita:2023rbz} oscillations. Our aim was to investigate the effects of hyperons and $\Delta$ baryons
on the non-radial $f$-mode~oscillations.

\par
This work is organized as follows. In~Section~\ref{sec:Model}, we give the details of the model considered. Subsequently, we describe the formulation of $f$-mode analysis in Section~\ref{sec:f-mode}, which discusses the stellar structure, Cowling approximation and tidal deformability. The~results obtained are discussed in Section~\ref{sec: results}. Finally, we provide a conclusion in Section~\ref{sec:conclusion}. 
Throughout this paper, we choose units such
that $G=\hbar=c=1$, where $G$ is the universal gravitational constant, $\hbar$ is the reduced Planck constant and $c$ is the speed of~light.

\section{Model}\label{sec:Model}
The effective Lagrangian introduced in Ref.~\cite{Malik2017} was applied here, in~which the nucleon isospin doublet $\psi_B$ interacts through the exchange of the pseudo-scalar meson $\pi$, the~scalar meson $\sigma$, the~iso-vector meson $\rho$ and the vector meson $\omega$:  
    {
    \begin{align*}
        {\cal{L}} &=  \bar{\psi}_{B}\Big[\left(i \gamma{\mu}\partial^{\mu}
        - g_{\omega B}\gamma_{\mu}\omega^{\mu} - \frac{1}{2}g_{\rho B}\vec{\rho_{\mu}}\cdot
        \vec{\tau}\gamma^{\mu}\right) - g_{\sigma B} (\sigma + i \gamma_{5}\vec{\tau}\cdot\vec{\pi})\Big]\psi_{B} \\
        & + \frac{1}{2}( \partial_{\mu}\vec{\pi}\cdot\partial^{\mu}\vec{\pi}  + \partial_{\mu}\sigma \partial^{\mu}\sigma )  - \frac{\lambda}{4}\left(x^{2}-x_{0}^{2}\right)^{2}- \frac{\lambda b} 
        {6 m_B^{2}}(x^{2}-x_{0}^{2})^{3}  \\ 
        & -  \frac{\lambda c}{8 m_B^{4}}(x^{2}-x_{0}^{2})^{4}- \frac{1}{4} F_{\mu\nu}F^{\mu\nu} + \frac{1}{2}g_{ \omega B}^{2}x^{2}\left(\omega_{\mu}
        \omega^{\mu}\right) \nonumber \\ 
        & - \frac{1}{4} \vec{R_{\mu \nu}}\cdot\vec{R^{\mu \nu}} + \frac{1}{2} 
        {m_\rho^\prime}^2\vec{\rho_{\mu}}\cdot\vec{\rho^{\mu}} + \eta_1 \left(\frac{1}{2} g_{\rho B}^2 x^2 \vec{\rho_{\mu}}\cdot\vec{\rho^{\mu}} \right).
    \end{align*}
    }%
\normalsize 

Following the interaction terms of the isospin doublet $\psi_B$, we have the kinetic and the non-linear terms in the pion field $\pi$ and the~scalar field $\sigma$ and~the higher order terms of the scalar field in terms of $x^2=\pi^2 + \sigma^2$. The~last two lines include the field strength and the mass term for the fields $\omega$ and $\rho$. The~final term includes the effects of cross-coupling between the $\sigma$ and $\rho$ mesons, and the coupling strength is denoted by $\eta_1$. We consider only the normal non-pion condensed state of matter and hence we take $< \pi>=0$ and $m_{\pi} = 0$.
The scalar field attains the vacuum expectation value $x_0$ due to the spontaneous symmetry breaking (SSB) of the chiral symmetry. {The masses of the baryon ($m_B$), the~scalar meson ($m_\sigma$) and the vector meson ($m_\omega$) are then given by
\begin{eqnarray}
m_B= g_{\sigma B} x_0,~ m_{\sigma} = \sqrt{2\lambda} x_0,~
m_{\omega} = g_{\omega B} x_0\ ,
\end{eqnarray}}%
where $\lambda=(m_\sigma^2-m_\pi^2)/2f_\pi^2$, with~$f_\pi$ being the pion decay constant which reflects the strength of the SSB. Due to the cross-coupling between the $\sigma$ and $\rho$ mesons,
the mass of the $\rho$ meson gets modified by the vacuum expectation value of the $\sigma$ meson as 
{$m_\rho^2=m^{\prime 2}_\rho+ \eta_1g_{\rho B}^2 x_0^2$}.

\par
This model was earlier applied to study matter at a finite temperature and low density, and~also the structure and composition of neutron stars \citep{Jha2008, Jha2008a}. The~details of the model considered in this work can be found in~\cite{Jyothilakshmi:2022hys}, where its role in the suppression of $r$-mode oscillations was also studied in detail. 
As mentioned before, here we studied two different models, one that incorporates the octet of baryons ($B = n^0, p^+, \Lambda^0, \Sigma^{+,0,-}, \Xi^{-,0}$) and another that includes delta baryons ($n^0, p^+, \Delta^{++,+,0,-}$), within~the same parameterization.
The expressions for the energy density and the pressure of the models considered are given by \vspace{6pt}
{
\begin{eqnarray}
\label{eosen}
      \epsilon &=& \frac{1}{\pi^2}\sum_{B}\int_0^{k_{F_B}}k^2\sqrt{k^2+m_B^*{^2}}dk
                       + \frac{m_B^2}{8 C_{\sigma B}}(1-Y^2)^2 \nonumber\\
      &&- \frac{b}{12 C_{\sigma B}C_{\omega B}}(1-Y^2)^3 + \frac{c}{16 m_B^2 C_{\sigma B}
            C_{\omega B}^2}(1-Y^2)^4  
         \nonumber \\
      &&+ \frac{1}{2}m_\rho^2\Big[1- \eta_1(1-Y^2) (C_{\rho B}/C_{\omega B})\Big] (\rho_{3B}^0)^2 \nonumber \\
      &&+ \frac{1}{2}m_{\omega}^2\omega_{0}^2Y^2 + \frac{1}{\pi^2}\sum_{L}\int_0^{k{F_L}}k^2\sqrt{k^2+m_L{^2}}dk,\\
\label{eospr}
   p &=& \frac{1}{3 \pi^2}\sum_{B}\int_0^{k_{F_B}}\frac{k^4}{\sqrt{k^2+m_B^*{^2}}}dk
         - \frac{m_B^2}{8 C_{\sigma B}}(1-Y^2)^2 \nonumber \\
     &&+ \frac{b}{12 C_{\sigma B}C_{\omega B}}(1-Y^2)^3 - \frac{c}{16 m_B^2
            C_{\sigma B}C_{\omega B}^2}(1-Y^2)^4     \nonumber\\
     && + \frac{1}{2}m_\rho^2 \Big[1- \eta_1(1-Y^2) (C_{\rho B}/C_{\omega B})\Big] (\rho_{3B}^0)^2
     \nonumber \\
     && + \frac{1}{2}m_{\omega}^2\omega_{0}^2Y^2 + \frac{1}{3 \pi^2}\sum_{L}\int_0^{k{F_L}}\frac{k^4}{\sqrt{k^2+m_L{^2}}}dk.
\end{eqnarray}}%
{Here, $k_{FB}$ is the fermi momentum of baryons,  and~$Y$ is a dimensionless variable defined as $x/x_0$. Also, $ C_{\sigma B}$, $ C_{\omega B}$ and $C_{\rho B}$, for a given baryon, $B$, are the coupling constants for $\sigma$, $\omega$ and $\rho$ respectively. }
The subscripts $B$ and $L$ denote the baryons and leptons under consideration, respectively. To~obtain the EoS, the stellar matter must satisfy the conditions of charge neutrality and chemical equilibrium. The~model parameters are listed in Table~\ref{saturation properties}, along with their saturation~properties.

\begin{table}[H]
  %  \centering
    \caption{Parameters 
 of the present model, such as the scalar, vector and iso-vector nucleon--meson coupling and its higher order interaction terms (top two rows), and the respective saturation properties (bottom two rows) are~displayed.}  \label{saturation properties}
    \setlength{\tabcolsep}{6.4mm}
    \begin{tabular}{cccccc}
    \toprule
         $C_{\sigma}$ & $C_{\omega}$ & $C_{\rho}$ & b & c & $\eta_{1}$  \\
         ($fm^{2}$) & ($fm^{2}$) & ($fm^{2}$) & ($fm^{2}$) & ($fm^{4}$) &  \\         
    \midrule
         8.81 & 2.16 & 13.00 & 12.08 & $-$36.47 & $-$0.85  \\
    \midrule
         E/A & ($m^{*}/m$) & K & $J_{0}$ & $L_{0}$ & $J_{1}$ \\
         (MeV) &  & (MeV) & (MeV) & (MeV) & (MeV) \\
    \midrule
         $-$16 & 0.84 & 210 & 32 & 60 & 24 \\
    \bottomrule     
    \end{tabular}
\end{table}
\unskip

%%%%%%%%%%%%%%%%%%%%%%%%%%%%%%%%%
%%%%%%%%%%%%%%%%%%%%%%%%%%%%%%%%%
\section{F-Mode~Analysis}\label{sec:f-mode}
\par
In this section, we describe the well-known Cowling approximation for non-radial oscillations along with the stellar structure equations and tidal deformability. The~line element for a static, spherically symmetric relativistic NS is given as
\begin{align}
   ds^{2}&=e^{2 \Phi(r)} dt^{2}-e^{2 \bar{\lambda}(r)} dr^{2}-r^{2}\left(d\theta ^{2}+\sin ^{2}\theta d\phi ^{2}\right).
\end{align}
Here, $\Phi(r)$ and $\bar{\lambda}(r)$ are the metric functions. The~solutions obtained by solving the Einstein Field equations for the given metric are the Tolman–Oppenheimer–Volkoff (TOV) equations~\cite{Tolman1939,Oppen1939}. They are as follows:
\begin{align}
     \label{tovP}
    \frac{dp}{dr}&=-\frac{\left[\epsilon+p\right ]
        \left[m+4\pi r^3 p\right ]}{r(r-2m)},\\
        \frac{dm}{dr}&= 4\pi r^2 \epsilon.
        \label{tovM}
\end{align}
\par
The TOV equations were integrated from the center to the surface of the NS to obtain the mass ($M$) and radius ($R$) of the star. Towards the center of the star, the~value of the pressure $p_c=p(r=0)=p(\rho_c)$ and $m_c=m(r=0)=0$, where $\rho_c$ is the central density. The~pressure tends to zero as it approaches the surface of the star. We obtained different configurations of the global properties (like $M$, $R$ and the compactness ($C=M/R$)) of the NS by repeating the integration for different central~densities. 

\par 
Next, we proceeded to obtain the $f$-mode frequencies using the Cowling approximation~\cite{Cowling:1941nqk}, as given in Ref.~\cite{PhysRevD.83.024014,Ranea-Sandoval:2018bgu,Flores:2013yqa}, for~which we made use of the solutions obtained from Equations~(\ref{tovP}) and (\ref{tovM}) to solve the following set of coupled differential equations:
\begin{align}
   \label{W_r} \frac{d W(r)}{dr}&=\frac{d \epsilon}{dp}\left[\omega^2r^2e^{\bar{\lambda} (r)-2\phi (r)}V (r)+\frac{d \Phi(r)}{dr} W (r)\right] \\
    &- l(l+1)e^{\bar{\lambda}(r)}V (r),\nonumber \\
  \label{V_r}  \frac{d V(r)}{dr} &= 2\frac{d\Phi (r)}{dr} V (r)-\frac{1}{r^2}e^{\bar{\lambda} (r)}W (r).
\end{align}
The functions $V (r)$ and $W (r)$, along with the frequency $\omega$, characterize the Lagrange displacement vector  ($\eta^i$) associated with perturbed fluid: 
\begin{equation*}
    \eta^{i}=\left ( e^{-\bar{\lambda} (r)}W (r),-V (r)\partial_{\theta},-V (r) \sin^{-2}{\theta}\  \partial {\phi}\right)\frac{Y_{lm} (\theta,\phi)}{r^2},
\end{equation*}
where $Y_{lm} (\theta,\phi)$ are the spherical harmonics. The~differential Equations~(\ref{W_r}) and (\ref{V_r}) obey the following boundary conditions towards the center of the star:
\begin{align}\label{BC-1}
    W (r) = Ar^{l+1}, \> V (r) = -\frac{A}{l} r^l.
\end{align}
 Here, $A$ is an arbitrary constant, and $l$ can take values of 2, 3, 4, etc. We did not consider $l=1$ because dipole oscillations do not give rise to gravitational waves~\cite{1991RSPSA.434..449C}. The~above differential equations were solved with some initial guess for $\omega ^2$. The~eigen frequency $\omega$ should satisfy the boundary condition at the surface given below:
\begin{equation} \label{BC-2}
    \omega^2e^{\bar{\lambda}(R)-2\Phi (R)}V (R)+\frac{1}{R^2}\frac{d\Phi (r)}{dr}\Big|_{r=R}W (R)=0.
\end{equation}
Equations~(\ref{W_r}) and (\ref{V_r}) were integrated from the center ($r=0$) to the surface ($r=R$), such that the boundary condition above was met.
After each integration, the initial guess of $\omega^2$ was modified and the calculations were repeated. 
Here, we used Ridders's method to obtain the $f$-mode or eigen frequency.
\par
The tidal deformability $\Lambda$ depends on the EoS through both the neutron star radius $R$ and the dimensionless Love number $k_2$ as $\Lambda = 2k_2R^5/3$. To~calculate $k_2$ for our EoS, we used the equations described in~\cite{Hinderer2008,Hind2010}. For~a static, spherically symmetric star in a static external quadrupolar tidal field, $\mathcal{E}_{ij}$, the~tidal deformability of a linear order is defined as $Q_{ij} = -\Lambda\mathcal{E}_{ij},$ where $Q_{ij}$ is the star's induced quadrupole moment. The~$l=2$ tidal Love number $k_2$ in terms of $\Lambda$ and $R$ is given as $k_2 = (3/2)\Lambda R^{-5}$. To~calculate $\Lambda$, we solved the following set of coupled first-order differential equations~\cite{Hind2010}\vspace{6pt}:
\begin{eqnarray}
\frac{dH}{dr}&=& \beta,\\
\frac{d\beta}{dr}&=&2 \left(1 - 2\frac{m}{r}\right)^{-1} H\left\{-2\pi
  \left[5\epsilon+9
    p+\mathcal{A}(\epsilon+p)\right]\phantom{\frac{3}{r^2}} \right. \nonumber\\
& + & \left. \quad\frac{3}{r^2}+2\left(1 - 2\frac{m}{r}\right)^{-1}
  \left(\frac{m}{r^2}+4\pi r p\right)^2\right\}\nonumber\\
&+&\frac{2\beta}{r}\left(1 -
  2\frac{m}{r}\right)^{-1}\left\{-1+\frac{m}{r}+2\pi r^2
  (\epsilon-p)\right\}.%\nonumber
\end{eqnarray}
Here, $\beta(r)=dH/dr$ and $\mathcal{A}=d\epsilon/dp$. The~above equations were combined with Equations~(\ref{tovP}) and  (\ref{tovM}) and solved simultaneously. The~integration is performed outwards starting at the values $H(r) = a_0r^2$ and $\beta(r) = 2a_0r$ as $r \to 0$. The~constant $a_0$ cancels out in the expression of the Love number and can therefore be chosen arbitrarily. It measures the deformation of the star. Since the stress--energy tensor $T_{\mu\nu} = 0$ outside the star, $H(r)$ is given in terms of the associated Legendre functions $Q_2^2\left(r/m-1\right) \sim r^{-3}$ and $P_2^2\left(r/m-1\right) \sim r^{2}$ at a large $r$. The~interior and exterior solutions are matched at $r=R$ to obtain a unique solution. Defining the following quantity, $y = { R \beta(R)}/{H(R)}$, for the solution at the interior, the~$l=2$ Love number is given as~\cite{Hind2010}
\begin{eqnarray}
k_2 &=& \frac{8C^5}{5}(1-2C)^2[2+2C(y-1)-y]\\
&\times&\bigg\{2C[6-3y+3C(5y-8)]\nonumber\\
      & +&4C^3[13-11y+C(3y-2)+2C^2(1+y)]\nonumber\\
      &+&3(1-2C)^2[2-y+2C(y-1)] \ln(1-2C)\bigg\}^{-1},\nonumber
\label{eq:k2}
\end{eqnarray}
where $C=M/R$ denotes the compactness of the star. We used the prescriptions given above for the stellar structure, non-radial oscillations and tidal deformability to study the $f$-mode oscillations for both hyperon and $\Delta$ baryon~EoSs.

\section{Results and Discussions}\label{sec: results} 
As mentioned before, we considered charge-neutral neutron star matter populated with hyperons and another EoS with $\Delta$ baryons, along with nucleons and leptons. In~the present work, we chose a scalar coupling, and~while keeping it alike for different hyperon species, we tuned the vector counterpart so as to reproduce the respective hyperon potentials and analyze the resulting effect on the EoS and the neutron star properties. The~hyperon couplings were defined with respect to the nucleon, such as the scalar coupling $x_{\sigma H} = \frac{g_{\sigma H}}{g_{\sigma N}}$, the~vector coupling $x_{\omega H} = \frac{g_{\omega H}}{g_{\omega N}}$ and the iso-vector coupling $x_{\rho H} = \frac{g_{\rho H}}{g_{\rho N}}$ for any hyperon species, $H$. The~binding energy of the hyperon species in symmetric nuclear matter could then be reproduced by the equation
\begin{equation}
    \left(\frac{B}{A}\right)_H = x_{\omega H}g_{\omega N} \omega_0 + m_H^* - m_H.
\end{equation}
In the equation above, $m_H^* = m_H \times Y$ is the effective mass of a particular hyperon species in matter. Recent experimental data indicate that the $\Lambda$ and $\Xi$ hyperons are bound with energies of $-28$ MeV and $-14$ MeV, respectively~\cite{Gomes2015, Gal2000}. It may also be noted that a very recent study reported the potential depth of $\Xi$ to be $\approx$$-20$ MeV~\cite{FRIEDMAN2021136555}. Therefore, in the present work, we set $U_\Xi = -20$ MeV. Also, we fixed the $\Sigma$ coupling by choosing the potential depth $U_{\Sigma} = 30$ MeV~\cite{Gal2000}. Throughout our calculation, we fixed $x_{\rho H} = x_{\omega H}$. In~the absence of any experimental data on the $\Delta$ potential, we fixed the corresponding couplings $x_{\sigma \Delta} = x_{\rho \Delta} = x_{\omega \Delta} = 1.2$~\cite{Raduta:2021xiz}, similarly to the ratio of the mass of $\Delta$ to~nucleons.

\begin{figure}[H]
  %\centering
  \subfloat[][\centering]{\includegraphics[width=0.48\linewidth,height = 5cm]{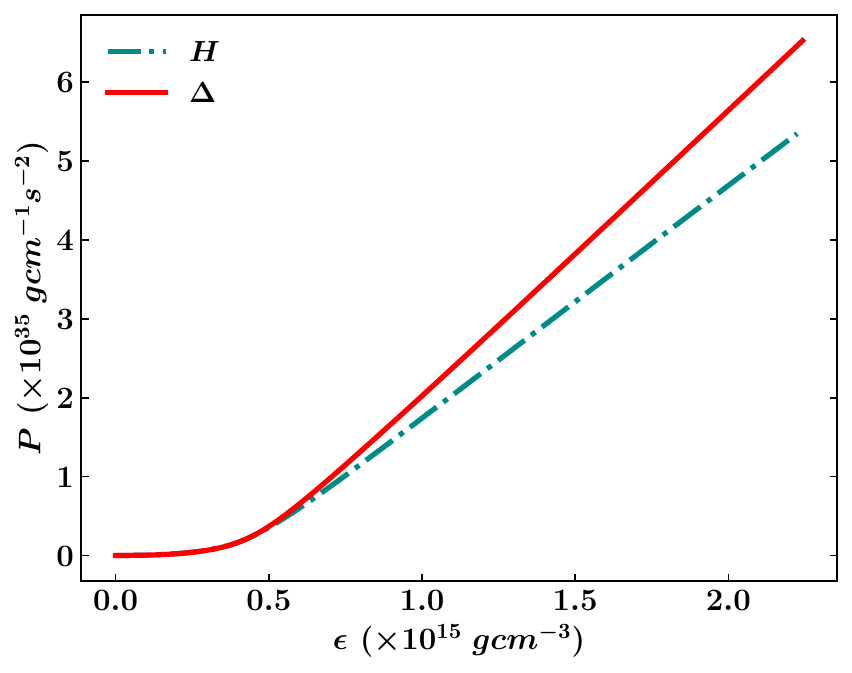} \label{Fig: EoS}} \quad
  \subfloat[][\centering]{\includegraphics[width=0.48\linewidth,height = 5cm]{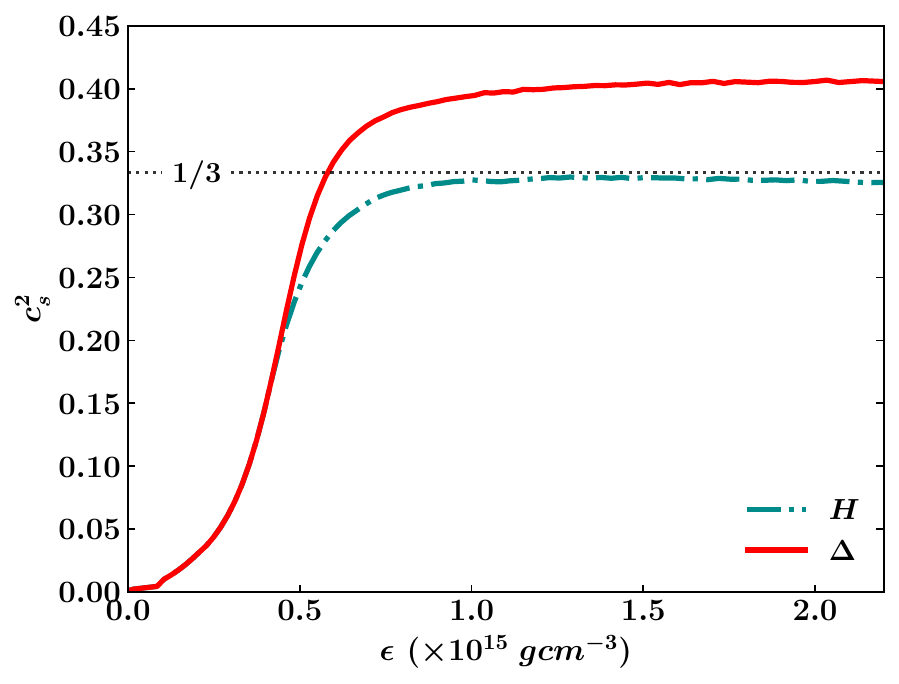} \label{Fig: sound}}
  \caption{(\textbf{a}) Hyperonic 
 ($H$) and delta ($\Delta$) equations of state of the models considered and (\textbf{b}) the square of the speed of sound as a function of the energy density for the EoSs. The~conformal limit $c_s^2=1/3$ is given by the dotted gray~line.}\label{fig1}
\end{figure}

The energy density versus the pressure of the models considered here is plotted in Figure~\ref{fig1}a. It can be seen that the softening effect of the EoS was much prominent in the case of hyperon-rich matter. % (denoted by $H$). 
This is understandable because most of the hyperons which appear in the neutron star matter happen to be lighter than the $\Delta$ baryons and hence appear early, as well as constituting an appreciable percentage of the neutron star matter. This is in contrast to the case of the $\Delta\,(1232)$ baryons, for~which the only species that appeared substantially in the charge-neutral star matter was $\Delta^{-}$. This underlying difference in the EoS and the composition of the neutron star matter must be reflected in the related global properties of the neutron star, which we discuss~next. \vspace{-12pt}

\par
We plotted the square of the speed of sound ($c_s^2=d p/d \epsilon$) as a function of the energy density $\epsilon$ in Figure~\ref{fig1}b. The~speed of sound is a measure of the stiffness of stellar matter, as evident from the plot, which shows that the hyperon-rich star matter was softer than the matter with $\Delta$ baryons. %(denoted by $\Delta$). 
At roughly $2\rho_{0}$, where $\rho_{0}$ is the normal nuclear density, the~exotic particles started to appear, and therefore the softening effect on the EoS could be seen at that particular density. Causality requires the speed of sound to satisfy the constraint $c_s^2 \leq 1$, while it has to satisfy the condition $c_s^2 > 0$ for thermodynamic stability. Furthermore, the~perturbative quantum
chromodynamics (QCD) results for extreme-density matter assume an upper limit of $c_s^2=1/3$. The~EoS with $\Delta$ baryons did not satisfy the perturbative QCD results beyond $2\rho_{0}$. 
Recently, several studies~\cite{Bedaque:2014sqa,Moustakidis:2016sab,Tews_2017} have shown that the $2 M_\odot$ constraints require a speed of sound squared that is greater than the conformal limit ($c_s^2=1/3$), indicating that the matter within the NS is a strongly interacting system. Our model with $\Delta$ baryons resulted in a stiffer EoS. Therefore, we note that the value of $c_s^2$ depends greatly on the internal composition of the model~considered. 

\par
\begin{figure}[H]
\centering
    \includegraphics[width=0.48\linewidth,height = 5cm]
    {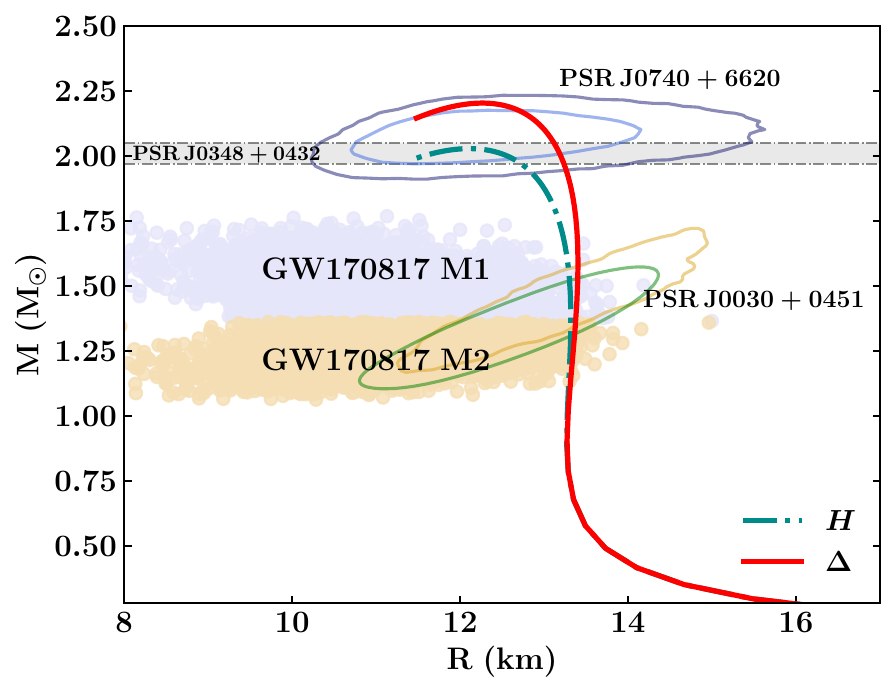}
    \caption{Maximum mass, $M$, versus radius, $R$, plot of the non-rotating hyperon $H$ and $\Delta$ baryon EoSs. The~horizontal band represent the observational limit of PSR J0348+0432 ($M = 2.01_{-0.04}^{+0.04} M_\odot$~\cite{Antoniadis2013}. Other limits from the NICER data of PSR J0740+6620 ($2.14_{-0.18}^{+0.20} M_\odot$ ($2$-$\sigma$ confidence interval)~\cite{NANOGrav:2019jur,Riley:2021pdl}) and PSR J0030+0451~\cite{Riley:2019yda,Miller:2019cac} are also given. The~observational constraints from the GW170817~\cite{LIGOScientific:2018cki} event are also shown.
    }
    \label{Fig:M-R}
    \end{figure}

 
We plotted the mass--radius profiles obtained by numerically solving the TOV equations given by Equations~(\ref{tovP}) and (\ref{tovM}) from the center to the surface of the star for the $H$ and $\Delta$ EoSs, shown in Figure~\ref{Fig:M-R}.
We obtained the maximum mass (and corresponding radius) for the $H$ and $\Delta$ EoSs as  $2.03 M_\odot$ ($12.11$ km) and $2.20 M_\odot$ ($12.26$ km), respectively. The~corresponding central densities were ($1.89$ and $1.84$) $\times 10^{15}$ g cm$^{-3}$, respectively. 
We noted that the $\Delta$ baryon matter resulted in a stiffer equation of state, which thereby gave a higher value for the maximum mass. In~Figure~\ref{Fig:M-R}, we have also given the observational constraints from PSR J0348+0432 ($M = 2.01_{-0.04}^{+0.04} M_\odot$ ($1$-
$\sigma$ confidence interval))~\cite{Antoniadis2013} and the pulsar PSR J0740+6620 ($2.14_{-0.09}^{+0.10} M_\odot$ ($1$-$\sigma$ confidence interval; the $2$-$\sigma$ confidence interval is $2.14_{-0.18}^{+0.20} M_\odot$))~\cite{NANOGrav:2019jur,Riley:2021pdl}.  We found that the maximum masses of the $H$ and $\Delta$ EoSs satisfied the given observational constraints. Furthermore, the~radius corresponding to the canonical mass $M_{1.4}$ for both the $H$ and $\Delta$ EoSs was $13.38$ and $13.31$ km, respectively, which is in good agreement with the NICER data of the pulsar PSR J0030+0451~\cite{Riley:2019yda,Miller:2019cac} and GW$170817$ observations~\cite{LIGOScientific:2018cki}. 

\par
\begin{figure}[H]
 % \centering
  \subfloat[][\centering]{\includegraphics[width=0.48\linewidth,height = 5cm]{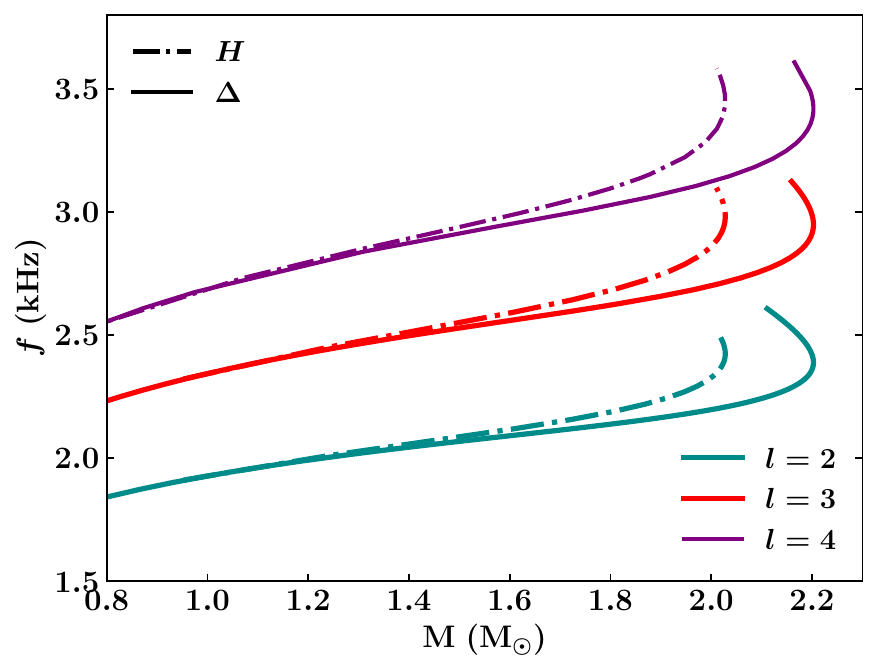} \label{Fig:M-f-l123}} \quad
   \subfloat[][\centering]{\includegraphics[width=0.48\linewidth,height = 5cm]{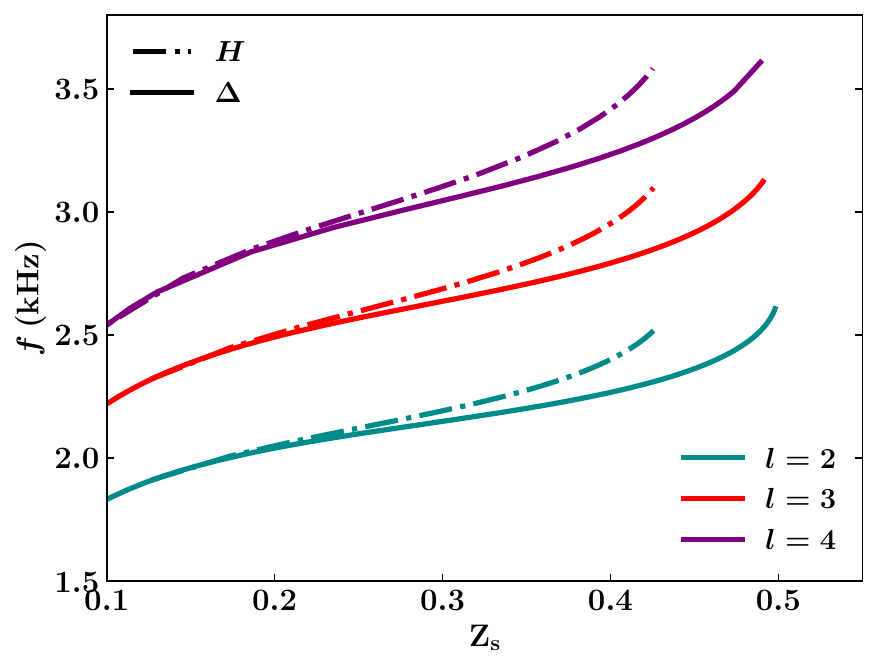} \label{f-Z}}
  \caption{$f$-mode frequencies as function of NS mass $M$ (\textbf{a
}) and redshift $Z_s$ (\textbf{b}) for $l=2,\;3$ and $4$ modes of~oscillation.}\label{fig3}
\end{figure}

 
Next, we proceeded to calculate the $f$-mode oscillation frequencies using Cowling approximation as discussed in Section~\ref{sec:f-mode}. We solved the differential equations given by Equations~(\ref{W_r}) and (\ref{V_r}) and repeated the integration for different values of $\omega$ until the condition given in Equation~(\ref{BC-2}) was satisfied. In~Figure~\ref{fig3}a, we show the plot of the $f$-mode frequencies as a function of the mass for the $l=2,\;3$ and $4$ modes for both the $H$ and $\Delta$ EoSs. We found that the $f$-mode frequencies of the $H$ and $\Delta$ EoSs showed a noticeable change in the higher mass region ($M>1.2M_\odot$), while the mode frequencies were similar in the lower mass region. 
The~$l=2$ mode frequencies ($f_{\text{max}}$) corresponding to the maximum mass for the $H$ and $\Delta$ baryon EoSs were $2.42$ and $2.39$ kHz. 
We found that $f_{\text{max}}$ was slightly decreased for the $\Delta$ EoS. The~$f_{\text{max}}$ values for $l=3\;(4)$ for the $H$ and  $\Delta$ EoSs were 2.94 (3.42) kHz and 2.98 (3.43) kHz, respectively. It is seen from Figure~\ref{fig3}a that the $f$-mode frequencies of both the EoSs increased with the mass until it attained a certain value. The~$f$-mode frequencies for $l= 2,\;3$ and $4$ for both $H$ and $\Delta$ lay approximately in the ranges of 1.84--2.42, 2.24--3 and 2.56--3.43 kHz, respectively, for a mass ranging from $0.8M_\odot$ to the maximum mass profile. We found that the frequencies of the $l=2$ mode obtained for the $\Delta$ EoS model lay within the range of mode frequencies obtained in Ref.~\cite{Kalita:2023rbz} with different $\Delta$ EoSs.
\par
Next, we obtained the values of compactness, $C=M/R$, and surface redshift, $Z_s = 1/\sqrt{1-2C}-1$, as a function of the $f$-mode frequency for the stellar models considered. 
The compactness of an NS quantifies how tightly its mass is packed within its radius, whereas the surface redshift measures how much light is redshifted as it escapes from the intense gravitational field of the star. The~$f$-mode frequencies as a function of $Z_s$ were plotted and are shown in Figure~\ref{fig3}b. The~values of compactness ($C_{\text{max}}$) and redshift ($Z_s^{\text{max}}$) corresponding to the maximum mass for the $H$ ($\Delta$) EoSs were $0.246\;(0.266)$ and $0.403\;(0.462)$, respectively. The~$C_{\text{max}}$ obtained for the $\Delta$ EoS was higher than that of the $H$ EoS due to the stiffness of the $\Delta$ EoS. Various observations have provided the values for redshift as $Z_s$ = 0.12--0.23 (E 1207.4-5209~\cite{Sanwal_2002}) and $Z_s = 0.205_{-0.003}^{+0.006}$ (RX J0720.4-3125~\cite{Hambaryan:2017wvm}). We found that the values of redshift obtained for the $H$ and $\Delta$ EoSs satisfied these observational~limits. \vspace{-12pt}
\par
   \begin{figure}[H]
 % \centering
  \subfloat[][\centering]{\includegraphics[width=0.48\linewidth,height = 5cm]{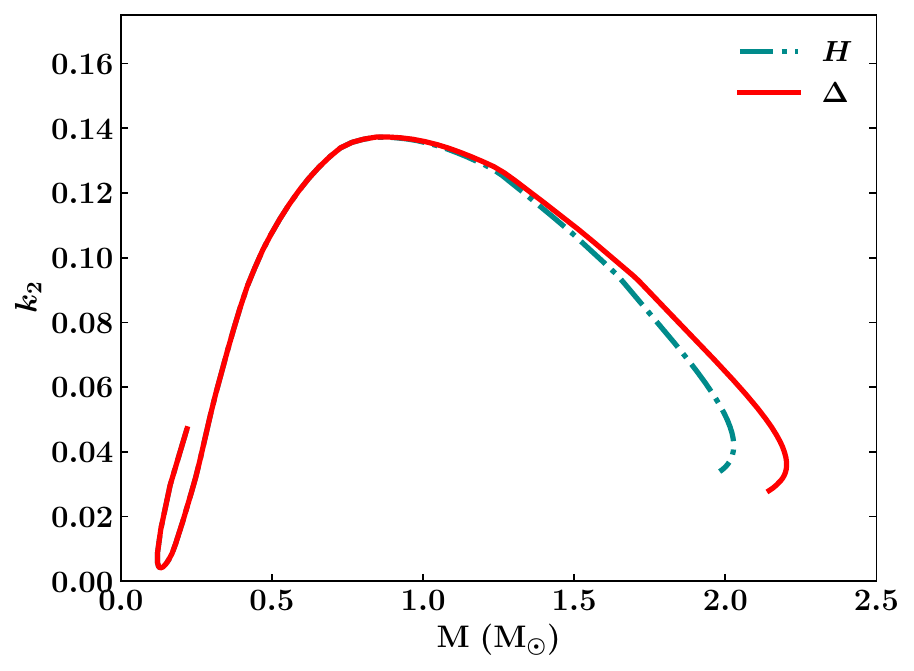}
  \label{Fig:k2-M}} \quad
  \subfloat[][\centering]{\includegraphics[width=0.48\linewidth,height = 5cm]{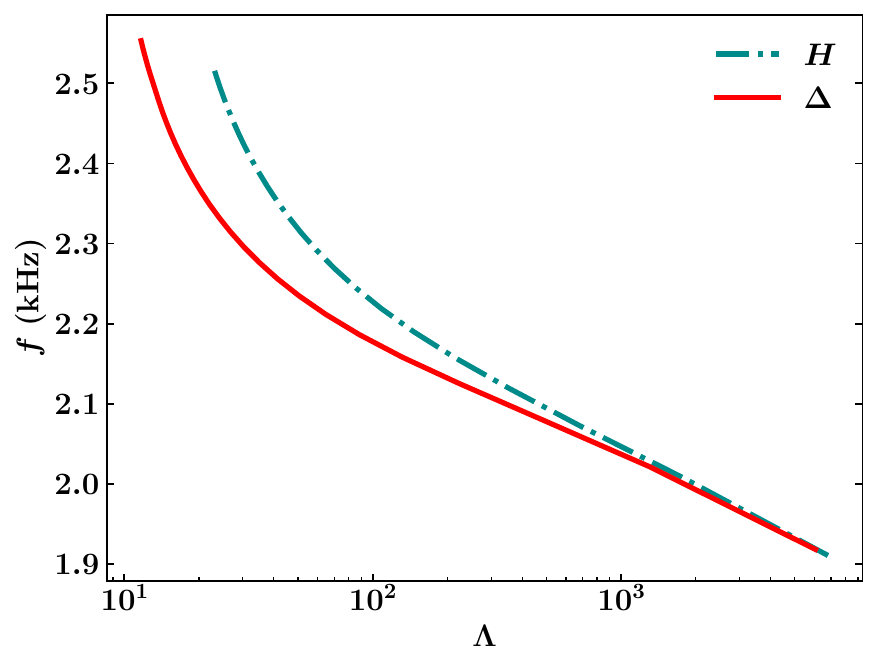} \label{Fig:L-F}}
  \caption{(\textbf{a}) The tidal Love number $k_2$ versus the stellar mass $M$ and (\textbf{b}) the $f$-mode frequencies, $f$, as a function of the tidal deformability $\Lambda$ for the EoSs~considered.}\label{fig4}
\end{figure}      

The tidal deformability provides key information about the neutron star matter. We calculated the dimensionless Love number ($k_2$) using Equation~(\ref{eq:k2}). In~Figure~\ref{fig4}a, we display a plot of $k_2$ versus the mass, $M$, of the stellar models considered. 
The value of $k_2$ remained the same for both of the EoSs considered until~it attained a maximum value of $\sim 0.137$, which had a corresponding mass value of $\sim$$0.9M_\odot$. Then, the value of $k_2$ decreased gradually and attained a value corresponding to the maximum mass ($k_2^{\text{max}}$) of $0.042$ and $0.036$ for the $H$ and $\Delta$ EoSs, respectively. The~tidal deformability ($\Lambda$) could be obtained from the Love number $k_2$ and the stellar radius ($R$) using the relation $k_2 = 2\Lambda R^{-5}/3$. We plotted the $f$-mode frequencies as a function of $\Lambda$ for the models considered, displayed in Figure~\ref{fig4}b. 
The $f$-mode frequencies as a function of $\Lambda$ for the $H$ and $\Delta$ EoSs were easily distinguishable, and in both cases the $\Lambda$ values decreased with an increase in the $f$-mode frequency until a certain value, and then they remained almost constant. The~values of $\Lambda$ corresponding to the maximum mass for the $H$ and $\Delta$ EoSs were $30.5$ and $17.8$, respectively. The~event GW190814 set a limit on the canonical tidal deformability of $\Lambda_{1.4}=616_{-158}^{+273}$ \cite{Abbott2020}. The~values of $\Lambda_{1.4}$ for the $H$ and $\Delta$ EoSs were around $\sim$$800$, which was in agreement with the constraint given by the GW event. Furthermore, in~Ref.~\cite{Wen:2019ouw}, the authors combined the constraints on the EoSs allowed by terrestrial nuclear experiments and tidal deformability data from GW170817. This gave them the limit for the $f$-mode frequency of a NS with $1.4M_\odot$ as 1.67--2.18 kHz. We noted that the $f$-mode frequencies corresponding to a $1.4M_\odot$ configuration for both the $H$ and $\Delta$ EoSs were approximately $\sim$$2$ kHz, which was in agreement with the given~limit.
 
\par
\begin{figure}[H]
   \centering
    \includegraphics[width=0.5\linewidth,height = 5cm]{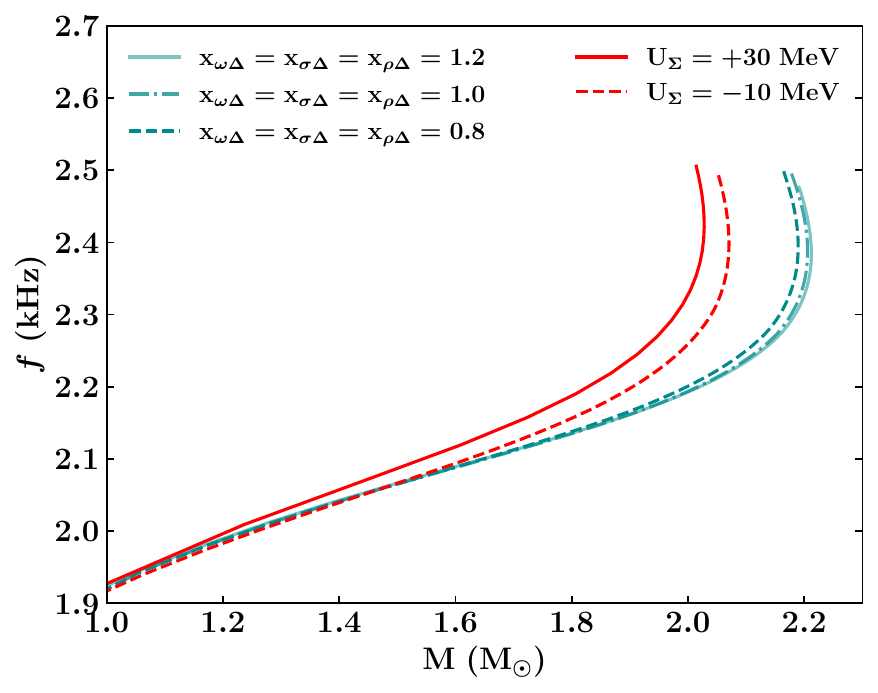}
    \caption{{The $l=2$ $f$-mode frequencies as a function of the NS mass $M$ for different EoSs obtained by varying the coupling constants and the potential depth of $\Sigma$ within the effective chiral model.}}
    \label{fig:f-M-1}
\end{figure}

{Next, we studied the $f$-mode oscillations by varying the potential depth $ U_\Sigma $ for $H$ models and the coupling constants for $\Delta$ baryons. In~Figure~\ref{fig:f-M-1}, we display the plot of the $f$-mode frequencies for the $ l=2 $ mode as a function of the NS mass for different parameterizations: \mbox{$ U_\Sigma = -10 $ MeV} and $ +30 $ MeV, and~$ x_{\sigma\Delta} = x_{\rho\Delta} = x_{\omega\Delta} = 0.8, 1.0, 1.2 $. 
We noted that the maximum mass increased when $ U_\Sigma $ was decreased from $ + 30 $ to $ -10 $ MeV. However, we found that the maximum mass showed only a negligible change when we varied the coupling constants of $\Delta$ baryons. We found that the range of $f$-mode frequencies did not change when varying the values of the $\Sigma$ potential and delta coupling constants. The~$f$-mode frequencies corresponding to the maximum mass ($ f_{\text{max}}$) did not change considerably when either the $\Sigma$ potentials or coupling constants were varied. 
}
\par
\begin{figure}[H]
   \centering
    
    \includegraphics[width=.5\linewidth,height = 5cm]
    {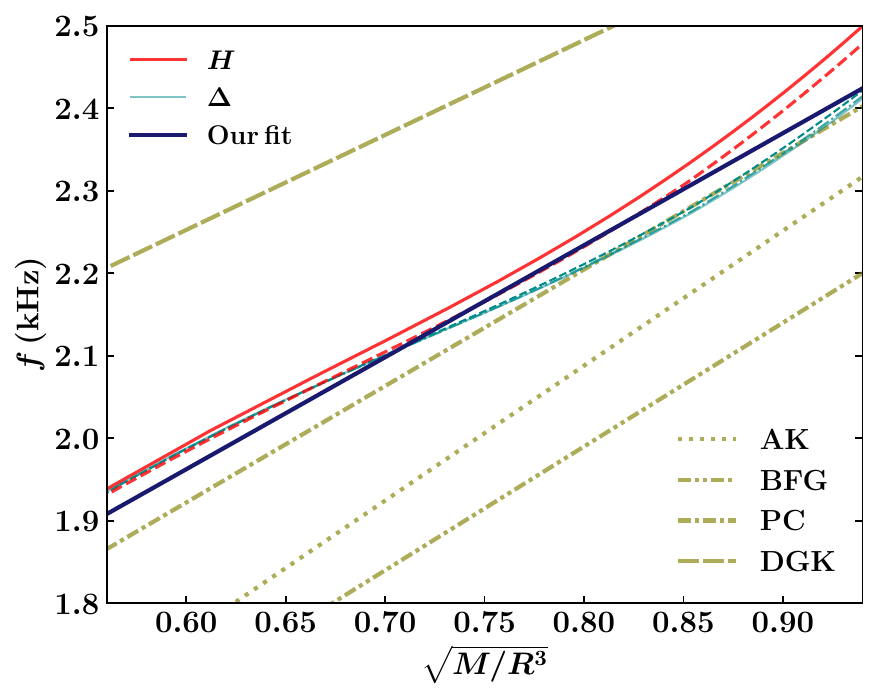}
    \caption{{$f$-mode frequency as a function of the average density for the EoSs considered, indicated by the same legend as in Figure~\ref{fig:f-M-1}.} The empirical relation obtained for our EoSs is given by the dark blue solid line. Other empirical relations taken from different works (N. Andersson $\&$ K. D. Kokkotas (1998) (AK)~\cite{PhysRevLett.77.4134},  O. Benhar~et~al. (2004) (BFG)~\cite{Benhar:2004xg}, D.Doneva~et~al. (2013) (DGK)~\cite{Doneva:2013zqa} and~B. K. Pradhan $\&$ D. Chatterjee (2021) (PC)~\cite{Pradhan:2020amo}) %, and~H. C. Das~et~al. (2021) (DKBP)~\cite{Das:2021dru})
     are given in an olive color.  }\label{Fig:Fit}
    \end{figure}

The fundamental idea behind gravitational wave asteroseismology is the same as that in traditional helioseismology, where stellar oscillations are studied to infer the properties of the interior of a star. 
Here, we proceeded to obtain a linear fit of the $f$-mode frequencies as a function of the average density, which was given as $f$ (kHz) $\approx$ $a+b\sqrt{\Tilde{M}/\Tilde{R}^3}$ %$\Tilde{\rho} = \sqrt{\Tilde{M}/\Tilde{R}^3}$ 
 (here, $\Tilde{M}=M/1.4M_\odot$ and $\Tilde{R}=R/10$ km) for our EoSs. In~Ref.~\cite{PhysRevLett.77.4134}, the~authors calculated the fitting relation using a polytropic EoS for the first time as $f$ (kHz) $\approx$ $0.17+2.3\sqrt{\Tilde{M}/\Tilde{R}^3}$. This relation was later modified by the authors for relativistic EoSs~\cite{Andersson:1997rn} to $f$ (kHz) $\approx$ $0.78+1.635\sqrt{\Tilde{M}/\Tilde{R}^3}$. Furthermore, in~Ref.~\cite{Benhar:2004xg}, the authors obtained empirical relations using hybrid EoSs. Recently, empirical relations using hyperonic and dark matter EoSs have been obtained in~\cite{Pradhan:2020amo}  and~\cite{Das:2021dru}, respectively. 
{We plotted the $f$-mode frequencies as a function of the average density for different parameterizations of the $H$ and $\Delta$ baryon models, shown in Figure~\ref{Fig:Fit}. We also calculated a linear fit between the frequency and average density for the models considered, which was obtained as $1.148$ kHz + $1.358\sqrt{\bar{M}/\bar{R}^3}$ kHz. }
 We compare the values of coefficients $a$ and $b$ obtained from different analyses in Table~\ref{tab:Fit}, and the fitting relations obtained are plotted in Figure~\ref{Fig:Fit}. The~empirical relations obtained in each case were different from one another, which was due to the difference in the composition of the matter considered inside the NS. We found that our data were closer to the fit obtained in Ref.~\cite{Pradhan:2020amo}. The~fitting relations can be used in order to constrain the EoS by measuring the $f$-mode frequency, which can be used to infer the mass and radius~values. 

\unskip
\unskip  
 \begin{table}[H]
\caption{Coefficients $a$ and $b$ of the fitting relation, $f$ (kHz) $\approx$ $a+b\sqrt{\Tilde{M}/\Tilde{R}^3}$, obtained from different works, for~the $l=2$ mode.\label{tab:Fit}}
\setlength{\tabcolsep}{8mm}
%\begin{center}
\begin{tabular}{ccc}  
\toprule
 \textbf{Works}  & \textbf{a (kHz)} & \textbf{b (kHz)}\\
\midrule
N. Andersson and K. D. Kokkotas (1998) \cite{PhysRevLett.77.4134} & 0.780 & 1.635\\
\midrule
O. Benhar~et~al. (2004) \cite{Benhar:2004xg} & 0.790&1.500\\
\midrule
D.Doneva~et~al. (2013) \cite{Doneva:2013zqa}&1.562 & 1.151 \\
\midrule
B. K. Pradhan and D. Chatterjee (2021) \cite{Pradhan:2020amo} &1.075 &1.412 \\
\midrule
{Our work} & {1.148} & {1.358}\\
\bottomrule
\end{tabular}
% \caption{ }
%\end{center}
\end{table}
\unskip

\section{Conclusions}\label{sec:conclusion}
In this work, we studied $f$-mode oscillations by employing Cowling approximation in neutron stars (NSs) with hyperons ($H$) and $\Delta$ baryons, using the effective chiral model with mesonic cross-coupling within the relativistic mean field theory. We found that the square of the speed of sound violated the conformal limit ($c_s^2=1/3$) for the $\Delta$ EoS, indicating a strongly interacting system. The~hyperonic system, on the other hand, gave a value of $c_s^2$, which was lower but closer to the conformal limit. We then obtained the static stellar properties of the models considered. We found that the maximum masses obtained for both the $H$ and $\Delta$ EoSs were consistent with the observational data of PSR J0348+0432, PSR J0740+6620 and~GW170817.  Then, we calculated the $f$-mode frequencies by employing Cowling approximation, which neglects the metric perturbations as a result of fluid oscillations. We calculated the $f$-mode frequency as a function of the stellar mass ($M$) and redshift ($Z_s$) for both the $H$ and $\Delta$ EoSs. %and tidal deformability $\Lambda$. %, which was compared with various observational datas available.
 We obtained the mode frequencies for masses ranging from $0.8M_\odot$ to the maximum mass configuration in the ranges of 1.84--2.42, 2.24--3 and 2.56--3.43 kHz for $l=2,\,3$ and $4$ modes, respectively. 
We found that the $f$-mode frequency ($f_{\text{max}}$) corresponding to the maximum mass was relatively higher for the stars with the $\Delta$ model.  %compared to the hyperon model. 
This was due to the greater stiffness of the $\Delta$ EoS compared to the hyperon EoS. 
The results obtained from our analysis were found to agree with the observational data from different pulsars and GW~events. 
\par 
We also studied the tidal properties of the models considered and it was found that the values of tidal deformability ($\Lambda$) for both the $H$ and $\Delta$ EoSs were in close agreement with the observational constraints obtained from the GW190814 event. We also obtained the $f$-mode frequencies as a function of $\Lambda$, and it was found that the mode frequency of a $1.4M_\odot$ star was approximately $\sim$$2$ kHz for both the $H$ and $\Delta$ NSs. We found that our results were in agreement with the limit provided in Ref.~\cite{Wen:2019ouw}. We hope that future observations from binary neutron star mergers could possibly provide more insights about the $f$-mode oscillations of NSs.
\par
{Next, we calculated the $f$-mode frequencies for the $l=2$ mode by varying the coupling constants and potential depth ($U_\Sigma$) for $\Delta$ baryons and hyperons, respectively. Our results show that the coupling constant had a negligible effect on both the frequencies and the mass of the stellar configuration. We also found that the ranges of the $f$-mode frequencies were almost the same for the models considered in our work.}
Further, we obtained an empirical relation (a relation obtained between the $f$-mode frequency and average density) for the set of EoSs considered. We also compared the relations obtained in other works with our work. We found that the empirical relations depend greatly on the internal composition of the~stars. 

\vspace{6pt}
\authorcontributions{Conceptualization, V.S. and T.K.J.; methodology, O.P.J. and P.E.S.K.; validation, O.P.J., P.E.S.K. and V.S.; formal analysis,  O.P.J. and P.E.S.K.; investigation, O.P.J., P.E.S.K., V.S., H.C. and T.K.J.; resources, O.P.J., P.E.S.K., V.S., H.C. and T.K.J.; data curation, O.P.J., P.E.S.K. and H.C.; writing---original draft preparation, O.P.J. and P.E.S.K.; writing---review and editing, V.S. and T.K.J.; visualization, V.S. and T.K.J.; supervision, V.S. and T.K.J.; funding acquisition, V.S. All authors have read and agreed to the published version of the manuscript.}

\funding{ A contribution towards the APC was given by the Provost Office, Amrita Vishwa Vidyapeetham. }

\dataavailability{Data is contained within the article. }
\acknowledgments{P.E.S.K. acknowledges the support of the Bonn-Cologne Graduate School of Physics and Astronomy (BCGS).} 

\conflictsofinterest{The authors declares no conflict of interest. }

\begin{adjustwidth}{-\extralength}{0cm}
\reftitle{References}
\PublishersNote{}
\end{adjustwidth}

%\bibliography{hyp}% Produces the bibliography via BibTeX.

\begin{thebibliography}{77}%
\makeatletter
\providecommand \@ifxundefined [1]{%
 \@ifx{#1\undefined}
}%
\providecommand \@ifnum [1]{%
 \ifnum #1\expandafter \@firstoftwo
 \else \expandafter \@secondoftwo
 \fi
}%
\providecommand \@ifx [1]{%
 \ifx #1\expandafter \@firstoftwo
 \else \expandafter \@secondoftwo
 \fi
}%
\providecommand \natexlab [1]{#1}%
\providecommand \enquote  [1]{``#1''}%
\providecommand \bibnamefont  [1]{#1}%
\providecommand \bibfnamefont [1]{#1}%
\providecommand \citenamefont [1]{#1}%
\providecommand \href@noop [0]{\@secondoftwo}%
\providecommand \href [0]{\begingroup \@sanitize@url \@href}%
\providecommand \@href[1]{\@@startlink{#1}\@@href}%
\providecommand \@@href[1]{\endgroup#1\@@endlink}%
\providecommand \@sanitize@url [0]{\catcode `\\12\catcode `\$12\catcode
  `\&12\catcode `\#12\catcode `\^12\catcode `\_12\catcode `\%12\relax}%
\providecommand \@@startlink[1]{}%
\providecommand \@@endlink[0]{}%
\providecommand \url  [0]{\begingroup\@sanitize@url \@url }%
\providecommand \@url [1]{\endgroup\@href {#1}{\urlprefix }}%
\providecommand \urlprefix  [0]{URL }%
\providecommand \Eprint [0]{\href }%
\providecommand \doibase [0]{http://dx.doi.org/}%
\providecommand \selectlanguage [0]{\@gobble}%
\providecommand \bibinfo  [0]{\@secondoftwo}%
\providecommand \bibfield  [0]{\@secondoftwo}%
\providecommand \translation [1]{[#1]}%
\providecommand \BibitemOpen [0]{}%
\providecommand \bibitemStop [0]{}%
\providecommand \bibitemNoStop [0]{.\EOS\space}%
\providecommand \EOS [0]{\spacefactor3000\relax}%
\providecommand \BibitemShut  [1]{\csname bibitem#1\endcsname}%
\let\auto@bib@innerbib\@empty
%</preamble>
\bibitem [{\citenamefont {Abbott}\ \emph {et~al.}(2017)\citenamefont {Abbott}
  \emph {et~al.}}]{Abbott2017}%
  \BibitemOpen
  \bibfield  {author} {\bibinfo {author} {\bibfnamefont {B.~P.}\ \bibnamefont
  {Abbott}} \emph {et~al.} (\bibinfo {collaboration} {LIGO Scientific,
  Virgo}),\ }\href {\doibase 10.1103/PhysRevLett.119.161101} {\bibfield
  {journal} {\bibinfo  {journal} {Phys. Rev. Lett.}\ }\textbf {\bibinfo
  {volume} {119}},\ \bibinfo {pages} {161101} (\bibinfo {year} {2017})},\
  \Eprint {http://arxiv.org/abs/1710.05832} {arXiv:1710.05832 [gr-qc]}
  \BibitemShut {NoStop}%
%%CITATION = ARXIV:1710.05832;%%
\bibitem [{\citenamefont {Abbott}\ \emph {et~al.}(2018)\citenamefont {Abbott}
  \emph {et~al.}}]{LIGOScientific:2018cki}%
  \BibitemOpen
  \bibfield  {author} {\bibinfo {author} {\bibfnamefont {B.~P.}\ \bibnamefont
  {Abbott}} \emph {et~al.} (\bibinfo {collaboration} {LIGO Scientific,
  Virgo}),\ }\href {\doibase 10.1103/PhysRevLett.121.161101} {\bibfield
  {journal} {\bibinfo  {journal} {Phys. Rev. Lett.}\ }\textbf {\bibinfo
  {volume} {121}},\ \bibinfo {pages} {161101} (\bibinfo {year} {2018})},\
  \Eprint {http://arxiv.org/abs/1805.11581} {arXiv:1805.11581 [gr-qc]}
  \BibitemShut {NoStop}%
\bibitem [{\citenamefont {Abbott}\ \emph
  {et~al.}(2020{\natexlab{a}})\citenamefont {Abbott} \emph
  {et~al.}}]{Abbott2020}%
  \BibitemOpen
  \bibfield  {author} {\bibinfo {author} {\bibfnamefont {R.}~\bibnamefont
  {Abbott}} \emph {et~al.} (\bibinfo {collaboration} {LIGO Scientific,
  Virgo}),\ }\href {\doibase 10.3847/2041-8213/ab960f} {\bibfield  {journal}
  {\bibinfo  {journal} {Astrophys. J.}\ }\textbf {\bibinfo {volume} {896}},\
  \bibinfo {pages} {L44} (\bibinfo {year} {2020}{\natexlab{a}})},\ \Eprint
  {http://arxiv.org/abs/2006.12611} {arXiv:2006.12611 [astro-ph.HE]}
  \BibitemShut {NoStop}%
%%CITATION = ARXIV:2006.12611;%%
\bibitem [{\citenamefont {Antoniadis}\ \emph {et~al.}(2013)\citenamefont
  {Antoniadis} \emph {et~al.}}]{Antoniadis2013}%
  \BibitemOpen
  \bibfield  {author} {\bibinfo {author} {\bibfnamefont {J.}~\bibnamefont
  {Antoniadis}} \emph {et~al.},\ }\href {\doibase 10.1126/science.1233232}
  {\bibfield  {journal} {\bibinfo  {journal} {Science}\ }\textbf {\bibinfo
  {volume} {340}},\ \bibinfo {pages} {6131} (\bibinfo {year} {2013})},\ \Eprint
  {http://arxiv.org/abs/1304.6875} {arXiv:1304.6875 [astro-ph.HE]} \BibitemShut
  {NoStop}%
\bibitem [{\citenamefont {Fonseca}\ \emph {et~al.}(2016)\citenamefont {Fonseca}
  \emph {et~al.}}]{Fonseca2016}%
  \BibitemOpen
  \bibfield  {author} {\bibinfo {author} {\bibfnamefont {E.}~\bibnamefont
  {Fonseca}} \emph {et~al.},\ }\href {\doibase 10.3847/0004-637X/832/2/167}
  {\bibfield  {journal} {\bibinfo  {journal} {Astrophys. J.}\ }\textbf
  {\bibinfo {volume} {832}},\ \bibinfo {pages} {167} (\bibinfo {year}
  {2016})},\ \Eprint {http://arxiv.org/abs/1603.00545} {arXiv:1603.00545
  [astro-ph.HE]} \BibitemShut {NoStop}%
%%CITATION = ARXIV:1603.00545;%%
\bibitem [{\citenamefont {Cromartie}\ \emph {et~al.}(2019)\citenamefont
  {Cromartie} \emph {et~al.}}]{NANOGrav:2019jur}%
  \BibitemOpen
  \bibfield  {author} {\bibinfo {author} {\bibfnamefont {H.~T.}\ \bibnamefont
  {Cromartie}} \emph {et~al.} (\bibinfo {collaboration} {NANOGrav}),\ }\href
  {\doibase 10.1038/s41550-019-0880-2} {\bibfield  {journal} {\bibinfo
  {journal} {Nature Astron.}\ }\textbf {\bibinfo {volume} {4}},\ \bibinfo
  {pages} {72} (\bibinfo {year} {2019})},\ \Eprint
  {http://arxiv.org/abs/1904.06759} {arXiv:1904.06759 [astro-ph.HE]}
  \BibitemShut {NoStop}%
\bibitem [{\citenamefont {Chandrasekhar}(1970)}]{CFS}%
  \BibitemOpen
  \bibfield  {author} {\bibinfo {author} {\bibfnamefont {S.}~\bibnamefont
  {Chandrasekhar}},\ }\href {\doibase 10.1103/PhysRevLett.24.611} {\bibfield
  {journal} {\bibinfo  {journal} {Phys. Rev. Lett.}\ }\textbf {\bibinfo
  {volume} {24}},\ \bibinfo {pages} {611} (\bibinfo {year} {1970})}\BibitemShut
  {NoStop}%
\bibitem [{\citenamefont {{Friedman}}\ and\ \citenamefont
  {{Schutz}}(1978)}]{Friedman78}%
  \BibitemOpen
  \bibfield  {author} {\bibinfo {author} {\bibfnamefont {J.~L.}\ \bibnamefont
  {{Friedman}}}\ and\ \bibinfo {author} {\bibfnamefont {B.~F.}\ \bibnamefont
  {{Schutz}}},\ }\href {\doibase 10.1086/156143} {\bibfield  {journal}
  {\bibinfo  {journal} {\apj}\ }\textbf {\bibinfo {volume} {222}},\ \bibinfo
  {pages} {281} (\bibinfo {year} {1978})}\BibitemShut {NoStop}%
\bibitem [{\citenamefont {Nayyar}\ and\ \citenamefont
  {Owen}(2006)}]{Nayyar:2005th}%
  \BibitemOpen
  \bibfield  {author} {\bibinfo {author} {\bibfnamefont {M.}~\bibnamefont
  {Nayyar}}\ and\ \bibinfo {author} {\bibfnamefont {B.~J.}\ \bibnamefont
  {Owen}},\ }\href {\doibase 10.1103/PhysRevD.73.084001} {\bibfield  {journal}
  {\bibinfo  {journal} {Phys. Rev. D}\ }\textbf {\bibinfo {volume} {73}},\
  \bibinfo {pages} {084001} (\bibinfo {year} {2006})},\ \Eprint
  {http://arxiv.org/abs/astro-ph/0512041} {arXiv:astro-ph/0512041} \BibitemShut
  {NoStop}%
\bibitem [{\citenamefont {Passamonti}\ and\ \citenamefont
  {Glampedakis}(2012)}]{Passamonti:2011eh}%
  \BibitemOpen
  \bibfield  {author} {\bibinfo {author} {\bibfnamefont {A.}~\bibnamefont
  {Passamonti}}\ and\ \bibinfo {author} {\bibfnamefont {K.}~\bibnamefont
  {Glampedakis}},\ }\href {\doibase 10.1111/j.1365-2966.2012.20849.x}
  {\bibfield  {journal} {\bibinfo  {journal} {Mon. Not. Roy. Astron. Soc.}\
  }\textbf {\bibinfo {volume} {422}},\ \bibinfo {pages} {3327} (\bibinfo {year}
  {2012})},\ \Eprint {http://arxiv.org/abs/1112.3931} {arXiv:1112.3931
  [astro-ph.SR]} \BibitemShut {NoStop}%
\bibitem [{\citenamefont {Doneva}\ \emph {et~al.}(2013)\citenamefont {Doneva},
  \citenamefont {Gaertig}, \citenamefont {Kokkotas},\ and\ \citenamefont
  {Kr\"uger}}]{Doneva:2013zqa}%
  \BibitemOpen
  \bibfield  {author} {\bibinfo {author} {\bibfnamefont {D.~D.}\ \bibnamefont
  {Doneva}}, \bibinfo {author} {\bibfnamefont {E.}~\bibnamefont {Gaertig}},
  \bibinfo {author} {\bibfnamefont {K.~D.}\ \bibnamefont {Kokkotas}}, \ and\
  \bibinfo {author} {\bibfnamefont {C.}~\bibnamefont {Kr\"uger}},\ }\href
  {\doibase 10.1103/PhysRevD.88.044052} {\bibfield  {journal} {\bibinfo
  {journal} {Phys. Rev. D}\ }\textbf {\bibinfo {volume} {88}},\ \bibinfo
  {pages} {044052} (\bibinfo {year} {2013})},\ \Eprint
  {http://arxiv.org/abs/1305.7197} {arXiv:1305.7197 [astro-ph.SR]} \BibitemShut
  {NoStop}%
\bibitem [{\citenamefont {Gittins}\ and\ \citenamefont
  {Andersson}(2023)}]{Gittins:2022rxs}%
  \BibitemOpen
  \bibfield  {author} {\bibinfo {author} {\bibfnamefont {F.}~\bibnamefont
  {Gittins}}\ and\ \bibinfo {author} {\bibfnamefont {N.}~\bibnamefont
  {Andersson}},\ }\href {\doibase 10.1093/mnras/stad672} {\bibfield  {journal}
  {\bibinfo  {journal} {Mon. Not. Roy. Astron. Soc.}\ }\textbf {\bibinfo
  {volume} {521}},\ \bibinfo {pages} {3043} (\bibinfo {year} {2023})},\ \Eprint
  {http://arxiv.org/abs/2212.04892} {arXiv:2212.04892 [gr-qc]} \BibitemShut
  {NoStop}%
\bibitem [{\citenamefont {Jyothilakshmi}\ \emph {et~al.}(2022)\citenamefont
  {Jyothilakshmi}, \citenamefont {Krishnan}, \citenamefont {Thakur},
  \citenamefont {Sreekanth},\ and\ \citenamefont
  {Jha}}]{Jyothilakshmi:2022hys}%
  \BibitemOpen
  \bibfield  {author} {\bibinfo {author} {\bibfnamefont {O.~P.}\ \bibnamefont
  {Jyothilakshmi}}, \bibinfo {author} {\bibfnamefont {P.~E.~S.}\ \bibnamefont
  {Krishnan}}, \bibinfo {author} {\bibfnamefont {P.}~\bibnamefont {Thakur}},
  \bibinfo {author} {\bibfnamefont {V.}~\bibnamefont {Sreekanth}}, \ and\
  \bibinfo {author} {\bibfnamefont {T.~K.}\ \bibnamefont {Jha}},\ }\href
  {\doibase 10.1093/mnras/stac2360} {\bibfield  {journal} {\bibinfo  {journal}
  {Mon. Not. Roy. Astron. Soc.}\ }\textbf {\bibinfo {volume} {516}},\ \bibinfo
  {pages} {3381} (\bibinfo {year} {2022})},\ \Eprint
  {http://arxiv.org/abs/2208.14436} {arXiv:2208.14436 [astro-ph.HE]}
  \BibitemShut {NoStop}%
\bibitem [{\citenamefont {Laskos-Patkos}\ and\ \citenamefont
  {Moustakidis}(2023)}]{Laskos-Patkos:2023cts}%
  \BibitemOpen
  \bibfield  {author} {\bibinfo {author} {\bibfnamefont {P.}~\bibnamefont
  {Laskos-Patkos}}\ and\ \bibinfo {author} {\bibfnamefont {C.~C.}\ \bibnamefont
  {Moustakidis}},\ }\href {\doibase 10.1103/PhysRevD.107.123023} {\bibfield
  {journal} {\bibinfo  {journal} {Phys. Rev. D}\ }\textbf {\bibinfo {volume}
  {107}},\ \bibinfo {pages} {123023} (\bibinfo {year} {2023})},\ \Eprint
  {http://arxiv.org/abs/2302.14537} {arXiv:2302.14537 [nucl-th]} \BibitemShut
  {NoStop}%
\bibitem [{\citenamefont {Passamonti}\ \emph {et~al.}(2013)\citenamefont
  {Passamonti}, \citenamefont {Gaertig},\ and\ \citenamefont
  {Kokkotas}}]{Passamonti2013}%
  \BibitemOpen
  \bibfield  {author} {\bibinfo {author} {\bibfnamefont {A.}~\bibnamefont
  {Passamonti}}, \bibinfo {author} {\bibfnamefont {E.}~\bibnamefont {Gaertig}},
  \ and\ \bibinfo {author} {\bibfnamefont {K.~D.}\ \bibnamefont {Kokkotas}},\
  }\href {\doibase 10.1088/1742-6596/453/1/012011} {\bibfield  {journal}
  {\bibinfo  {journal} {Journal of Physics: Conference Series}\ }\textbf
  {\bibinfo {volume} {453}},\ \bibinfo {pages} {012011} (\bibinfo {year}
  {2013})}\BibitemShut {NoStop}%
\bibitem [{\citenamefont {Lindblom}\ and\ \citenamefont
  {Detweiler}(1983)}]{Lindblom:1983ps}%
  \BibitemOpen
  \bibfield  {author} {\bibinfo {author} {\bibfnamefont {L.}~\bibnamefont
  {Lindblom}}\ and\ \bibinfo {author} {\bibfnamefont {S.~L.}\ \bibnamefont
  {Detweiler}},\ }\href {\doibase 10.1086/190884} {\bibfield  {journal}
  {\bibinfo  {journal} {Astrophys. J. Suppl.}\ }\textbf {\bibinfo {volume}
  {53}},\ \bibinfo {pages} {73} (\bibinfo {year} {1983})}\BibitemShut {NoStop}%
\bibitem [{\citenamefont {Cowling}(1941)}]{Cowling:1941nqk}%
  \BibitemOpen
  \bibfield  {author} {\bibinfo {author} {\bibfnamefont {T.~G.}\ \bibnamefont
  {Cowling}},\ }\href {\doibase 10.1093/mnras/101.8.367} {\bibfield  {journal}
  {\bibinfo  {journal} {Mon. Not. Roy. Astron. Soc.}\ }\textbf {\bibinfo
  {volume} {101}},\ \bibinfo {pages} {367} (\bibinfo {year}
  {1941})}\BibitemShut {NoStop}%
\bibitem [{\citenamefont {Sotani}\ and\ \citenamefont
  {Takiwaki}(2020)}]{Sotani:2020mwc}%
  \BibitemOpen
  \bibfield  {author} {\bibinfo {author} {\bibfnamefont {H.}~\bibnamefont
  {Sotani}}\ and\ \bibinfo {author} {\bibfnamefont {T.}~\bibnamefont
  {Takiwaki}},\ }\href {\doibase 10.1103/PhysRevD.102.063025} {\bibfield
  {journal} {\bibinfo  {journal} {Phys. Rev. D}\ }\textbf {\bibinfo {volume}
  {102}},\ \bibinfo {pages} {063025} (\bibinfo {year} {2020})},\ \Eprint
  {http://arxiv.org/abs/2009.05206} {arXiv:2009.05206 [astro-ph.HE]}
  \BibitemShut {NoStop}%
\bibitem [{\citenamefont {Sotani}\ \emph {et~al.}(2011)\citenamefont {Sotani},
  \citenamefont {Yasutake}, \citenamefont {Maruyama},\ and\ \citenamefont
  {Tatsumi}}]{PhysRevD.83.024014}%
  \BibitemOpen
  \bibfield  {author} {\bibinfo {author} {\bibfnamefont {H.}~\bibnamefont
  {Sotani}}, \bibinfo {author} {\bibfnamefont {N.}~\bibnamefont {Yasutake}},
  \bibinfo {author} {\bibfnamefont {T.}~\bibnamefont {Maruyama}}, \ and\
  \bibinfo {author} {\bibfnamefont {T.}~\bibnamefont {Tatsumi}},\ }\href
  {\doibase 10.1103/PhysRevD.83.024014} {\bibfield  {journal} {\bibinfo
  {journal} {Phys. Rev. D}\ }\textbf {\bibinfo {volume} {83}},\ \bibinfo
  {pages} {024014} (\bibinfo {year} {2011})}\BibitemShut {NoStop}%
\bibitem [{\citenamefont {Flores}\ and\ \citenamefont
  {Lugones}(2014)}]{Flores:2013yqa}%
  \BibitemOpen
  \bibfield  {author} {\bibinfo {author} {\bibfnamefont {C.~V.}\ \bibnamefont
  {Flores}}\ and\ \bibinfo {author} {\bibfnamefont {G.}~\bibnamefont
  {Lugones}},\ }\href {\doibase 10.1088/0264-9381/31/15/155002} {\bibfield
  {journal} {\bibinfo  {journal} {Class. Quant. Grav.}\ }\textbf {\bibinfo
  {volume} {31}},\ \bibinfo {pages} {155002} (\bibinfo {year} {2014})},\
  \Eprint {http://arxiv.org/abs/1310.0554} {arXiv:1310.0554 [astro-ph.HE]}
  \BibitemShut {NoStop}%
\bibitem [{\citenamefont {Ranea-Sandoval}\ \emph {et~al.}(2018)\citenamefont
  {Ranea-Sandoval}, \citenamefont {Guilera}, \citenamefont {Mariani},\ and\
  \citenamefont {Orsaria}}]{Ranea-Sandoval:2018bgu}%
  \BibitemOpen
  \bibfield  {author} {\bibinfo {author} {\bibfnamefont {I.~F.}\ \bibnamefont
  {Ranea-Sandoval}}, \bibinfo {author} {\bibfnamefont {O.~M.}\ \bibnamefont
  {Guilera}}, \bibinfo {author} {\bibfnamefont {M.}~\bibnamefont {Mariani}}, \
  and\ \bibinfo {author} {\bibfnamefont {M.~G.}\ \bibnamefont {Orsaria}},\
  }\href {\doibase 10.1088/1475-7516/2018/12/031} {\bibfield  {journal}
  {\bibinfo  {journal} {JCAP}\ }\textbf {\bibinfo {volume} {12}},\ \bibinfo
  {pages} {031} (\bibinfo {year} {2018})},\ \Eprint
  {http://arxiv.org/abs/1807.02166} {arXiv:1807.02166 [astro-ph.HE]}
  \BibitemShut {NoStop}%
\bibitem [{\citenamefont {V\'asquez~Flores}\ \emph {et~al.}(2019)\citenamefont
  {V\'asquez~Flores}, \citenamefont {Parisi}, \citenamefont {Chen},\ and\
  \citenamefont {Lugones}}]{VasquezFlores:2019eht}%
  \BibitemOpen
  \bibfield  {author} {\bibinfo {author} {\bibfnamefont {C.}~\bibnamefont
  {V\'asquez~Flores}}, \bibinfo {author} {\bibfnamefont {A.}~\bibnamefont
  {Parisi}}, \bibinfo {author} {\bibfnamefont {C.-S.}\ \bibnamefont {Chen}}, \
  and\ \bibinfo {author} {\bibfnamefont {G.}~\bibnamefont {Lugones}},\ }\href
  {\doibase 10.1088/1475-7516/2019/06/051} {\bibfield  {journal} {\bibinfo
  {journal} {JCAP}\ }\textbf {\bibinfo {volume} {06}},\ \bibinfo {pages} {051}
  (\bibinfo {year} {2019})},\ \Eprint {http://arxiv.org/abs/1901.07157}
  {arXiv:1901.07157 [hep-ph]} \BibitemShut {NoStop}%
\bibitem [{\citenamefont {Pradhan}\ and\ \citenamefont
  {Chatterjee}(2021)}]{Pradhan:2020amo}%
  \BibitemOpen
  \bibfield  {author} {\bibinfo {author} {\bibfnamefont {B.~K.}\ \bibnamefont
  {Pradhan}}\ and\ \bibinfo {author} {\bibfnamefont {D.}~\bibnamefont
  {Chatterjee}},\ }\href {\doibase 10.1103/PhysRevC.103.035810} {\bibfield
  {journal} {\bibinfo  {journal} {Phys. Rev. C}\ }\textbf {\bibinfo {volume}
  {103}},\ \bibinfo {pages} {035810} (\bibinfo {year} {2021})},\ \Eprint
  {http://arxiv.org/abs/2011.02204} {arXiv:2011.02204 [astro-ph.HE]}
  \BibitemShut {NoStop}%
\bibitem [{\citenamefont {Kumar}\ \emph {et~al.}(2023)\citenamefont {Kumar},
  \citenamefont {Mishra},\ and\ \citenamefont {Malik}}]{Kumar:2021hzo}%
  \BibitemOpen
  \bibfield  {author} {\bibinfo {author} {\bibfnamefont {D.}~\bibnamefont
  {Kumar}}, \bibinfo {author} {\bibfnamefont {H.}~\bibnamefont {Mishra}}, \
  and\ \bibinfo {author} {\bibfnamefont {T.}~\bibnamefont {Malik}},\ }\href
  {\doibase 10.1088/1475-7516/2023/02/015} {\bibfield  {journal} {\bibinfo
  {journal} {JCAP}\ }\textbf {\bibinfo {volume} {02}},\ \bibinfo {pages} {015}
  (\bibinfo {year} {2023})},\ \Eprint {http://arxiv.org/abs/2110.00324}
  {arXiv:2110.00324 [hep-ph]} \BibitemShut {NoStop}%
\bibitem [{\citenamefont {Das}\ \emph {et~al.}(2021)\citenamefont {Das},
  \citenamefont {Kumar}, \citenamefont {Biswal},\ and\ \citenamefont
  {Patra}}]{Das:2021dru}%
  \BibitemOpen
  \bibfield  {author} {\bibinfo {author} {\bibfnamefont {H.~C.}\ \bibnamefont
  {Das}}, \bibinfo {author} {\bibfnamefont {A.}~\bibnamefont {Kumar}}, \bibinfo
  {author} {\bibfnamefont {S.~K.}\ \bibnamefont {Biswal}}, \ and\ \bibinfo
  {author} {\bibfnamefont {S.~K.}\ \bibnamefont {Patra}},\ }\href {\doibase
  10.1103/PhysRevD.104.123006} {\bibfield  {journal} {\bibinfo  {journal}
  {Phys. Rev. D}\ }\textbf {\bibinfo {volume} {104}},\ \bibinfo {pages}
  {123006} (\bibinfo {year} {2021})},\ \Eprint
  {http://arxiv.org/abs/2109.01851} {arXiv:2109.01851 [nucl-th]} \BibitemShut
  {NoStop}%
\bibitem [{\citenamefont {Jyothilakshmi}\ \emph {et~al.}(2024)\citenamefont
  {Jyothilakshmi}, \citenamefont {Naik},\ and\ \citenamefont
  {Sreekanth}}]{Jyothilakshmi:2024zqn}%
  \BibitemOpen
  \bibfield  {author} {\bibinfo {author} {\bibfnamefont {O.~P.}\ \bibnamefont
  {Jyothilakshmi}}, \bibinfo {author} {\bibfnamefont {L.~J.}\ \bibnamefont
  {Naik}}, \ and\ \bibinfo {author} {\bibfnamefont {V.}~\bibnamefont
  {Sreekanth}},\ }\href {\doibase 10.1140/epjc/s10052-024-12776-9} {\bibfield
  {journal} {\bibinfo  {journal} {Eur. Phys. J. C}\ }\textbf {\bibinfo {volume}
  {84}},\ \bibinfo {pages} {427} (\bibinfo {year} {2024})},\ \Eprint
  {http://arxiv.org/abs/2403.00711} {arXiv:2403.00711 [gr-qc]} \BibitemShut
  {NoStop}%
\bibitem [{\citenamefont {Lattimer}\ and\ \citenamefont
  {Prakash}(2007)}]{Lattimer:2006xb}%
  \BibitemOpen
  \bibfield  {author} {\bibinfo {author} {\bibfnamefont {J.~M.}\ \bibnamefont
  {Lattimer}}\ and\ \bibinfo {author} {\bibfnamefont {M.}~\bibnamefont
  {Prakash}},\ }\href {\doibase 10.1016/j.physrep.2007.02.003} {\bibfield
  {journal} {\bibinfo  {journal} {Phys. Rept.}\ }\textbf {\bibinfo {volume}
  {442}},\ \bibinfo {pages} {109} (\bibinfo {year} {2007})},\ \Eprint
  {http://arxiv.org/abs/astro-ph/0612440} {arXiv:astro-ph/0612440} \BibitemShut
  {NoStop}%
\bibitem [{\citenamefont {Wen}\ \emph {et~al.}(2019)\citenamefont {Wen},
  \citenamefont {Li}, \citenamefont {Chen},\ and\ \citenamefont
  {Zhang}}]{Wen:2019ouw}%
  \BibitemOpen
  \bibfield  {author} {\bibinfo {author} {\bibfnamefont {D.-H.}\ \bibnamefont
  {Wen}}, \bibinfo {author} {\bibfnamefont {B.-A.}\ \bibnamefont {Li}},
  \bibinfo {author} {\bibfnamefont {H.-Y.}\ \bibnamefont {Chen}}, \ and\
  \bibinfo {author} {\bibfnamefont {N.-B.}\ \bibnamefont {Zhang}},\ }\href
  {\doibase 10.1103/PhysRevC.99.045806} {\bibfield  {journal} {\bibinfo
  {journal} {Phys. Rev. C}\ }\textbf {\bibinfo {volume} {99}},\ \bibinfo
  {pages} {045806} (\bibinfo {year} {2019})},\ \Eprint
  {http://arxiv.org/abs/1901.03779} {arXiv:1901.03779 [gr-qc]} \BibitemShut
  {NoStop}%
\bibitem [{\citenamefont {Gell-Mann}\ and\ \citenamefont
  {Levy}(1960)}]{Gell-Mann:1960mvl}%
  \BibitemOpen
  \bibfield  {author} {\bibinfo {author} {\bibfnamefont {M.}~\bibnamefont
  {Gell-Mann}}\ and\ \bibinfo {author} {\bibfnamefont {M.}~\bibnamefont
  {Levy}},\ }\href {\doibase 10.1007/BF02859738} {\bibfield  {journal}
  {\bibinfo  {journal} {Nuovo Cim.}\ }\textbf {\bibinfo {volume} {16}},\
  \bibinfo {pages} {705} (\bibinfo {year} {1960})}\BibitemShut {NoStop}%
\bibitem [{\citenamefont {Lee}\ and\ \citenamefont {Wick}(1974)}]{Lee:1974ma}%
  \BibitemOpen
  \bibfield  {author} {\bibinfo {author} {\bibfnamefont {T.~D.}\ \bibnamefont
  {Lee}}\ and\ \bibinfo {author} {\bibfnamefont {G.~C.}\ \bibnamefont {Wick}},\
  }\href {\doibase 10.1103/PhysRevD.9.2291} {\bibfield  {journal} {\bibinfo
  {journal} {Phys. Rev. D}\ }\textbf {\bibinfo {volume} {9}},\ \bibinfo {pages}
  {2291} (\bibinfo {year} {1974})}\BibitemShut {NoStop}%
\bibitem [{\citenamefont {Lee}\ and\ \citenamefont
  {Margulies}(1975)}]{Lee:1974uu}%
  \BibitemOpen
  \bibfield  {author} {\bibinfo {author} {\bibfnamefont {T.~D.}\ \bibnamefont
  {Lee}}\ and\ \bibinfo {author} {\bibfnamefont {M.}~\bibnamefont
  {Margulies}},\ }\href {\doibase 10.1103/PhysRevD.12.4008} {\bibfield
  {journal} {\bibinfo  {journal} {Phys. Rev. D}\ }\textbf {\bibinfo {volume}
  {11}},\ \bibinfo {pages} {1591} (\bibinfo {year} {1975})}\BibitemShut
  {NoStop}%
\bibitem [{\citenamefont {Furnstahl}\ and\ \citenamefont
  {Serot}(1993)}]{Furnstahl:1993wx}%
  \BibitemOpen
  \bibfield  {author} {\bibinfo {author} {\bibfnamefont {R.~J.}\ \bibnamefont
  {Furnstahl}}\ and\ \bibinfo {author} {\bibfnamefont {B.~D.}\ \bibnamefont
  {Serot}},\ }\href {\doibase 10.1016/0370-2693(93)90649-3} {\bibfield
  {journal} {\bibinfo  {journal} {Phys. Lett. B}\ }\textbf {\bibinfo {volume}
  {316}},\ \bibinfo {pages} {12} (\bibinfo {year} {1993})}\BibitemShut
  {NoStop}%
\bibitem [{\citenamefont {Serot}(2002)}]{Serot:2002ei}%
  \BibitemOpen
  \bibfield  {author} {\bibinfo {author} {\bibfnamefont {B.~D.}\ \bibnamefont
  {Serot}},\ }\href {\doibase 10.1142/S0217979203020284} {\bibfield  {journal}
  {\bibinfo  {journal} {Ser. Adv. Quant. Many Body Theor.}\ }\textbf {\bibinfo
  {volume} {6}},\ \bibinfo {pages} {207} (\bibinfo {year} {2002})},\ \Eprint
  {http://arxiv.org/abs/nucl-th/0201083} {arXiv:nucl-th/0201083} \BibitemShut
  {NoStop}%
\bibitem [{\citenamefont {Heide}\ \emph {et~al.}(1994)\citenamefont {Heide},
  \citenamefont {Rudaz},\ and\ \citenamefont {Ellis}}]{Heide:1993yz}%
  \BibitemOpen
  \bibfield  {author} {\bibinfo {author} {\bibfnamefont {E.~K.}\ \bibnamefont
  {Heide}}, \bibinfo {author} {\bibfnamefont {S.}~\bibnamefont {Rudaz}}, \ and\
  \bibinfo {author} {\bibfnamefont {P.~J.}\ \bibnamefont {Ellis}},\ }\href
  {\doibase 10.1016/0375-9474(94)90717-X} {\bibfield  {journal} {\bibinfo
  {journal} {Nucl. Phys. A}\ }\textbf {\bibinfo {volume} {571}},\ \bibinfo
  {pages} {713} (\bibinfo {year} {1994})},\ \Eprint
  {http://arxiv.org/abs/nucl-th/9308002} {arXiv:nucl-th/9308002} \BibitemShut
  {NoStop}%
\bibitem [{\citenamefont {Mishustin}\ \emph {et~al.}(1993)\citenamefont
  {Mishustin}, \citenamefont {Bondorf},\ and\ \citenamefont
  {Rho}}]{Mishustin:1993ub}%
  \BibitemOpen
  \bibfield  {author} {\bibinfo {author} {\bibfnamefont {I.}~\bibnamefont
  {Mishustin}}, \bibinfo {author} {\bibfnamefont {J.}~\bibnamefont {Bondorf}},
  \ and\ \bibinfo {author} {\bibfnamefont {M.}~\bibnamefont {Rho}},\ }\href
  {\doibase 10.1016/0375-9474(93)90319-S} {\bibfield  {journal} {\bibinfo
  {journal} {Nucl. Phys. A}\ }\textbf {\bibinfo {volume} {555}},\ \bibinfo
  {pages} {215} (\bibinfo {year} {1993})}\BibitemShut {NoStop}%
\bibitem [{\citenamefont {Papazoglou}\ \emph {et~al.}(1999)\citenamefont
  {Papazoglou}, \citenamefont {Zschiesche}, \citenamefont {Schramm},
  \citenamefont {Schaffner-Bielich}, \citenamefont {Stoecker},\ and\
  \citenamefont {Greiner}}]{Papazoglou:1998vr}%
  \BibitemOpen
  \bibfield  {author} {\bibinfo {author} {\bibfnamefont {P.}~\bibnamefont
  {Papazoglou}}, \bibinfo {author} {\bibfnamefont {D.}~\bibnamefont
  {Zschiesche}}, \bibinfo {author} {\bibfnamefont {S.}~\bibnamefont {Schramm}},
  \bibinfo {author} {\bibfnamefont {J.}~\bibnamefont {Schaffner-Bielich}},
  \bibinfo {author} {\bibfnamefont {H.}~\bibnamefont {Stoecker}}, \ and\
  \bibinfo {author} {\bibfnamefont {W.}~\bibnamefont {Greiner}},\ }\href
  {\doibase 10.1103/PhysRevC.59.411} {\bibfield  {journal} {\bibinfo  {journal}
  {Phys. Rev. C}\ }\textbf {\bibinfo {volume} {59}},\ \bibinfo {pages} {411}
  (\bibinfo {year} {1999})},\ \Eprint {http://arxiv.org/abs/nucl-th/9806087}
  {arXiv:nucl-th/9806087} \BibitemShut {NoStop}%
\bibitem [{\citenamefont {Schramm}(2002)}]{Schramm:2002xi}%
  \BibitemOpen
  \bibfield  {author} {\bibinfo {author} {\bibfnamefont {S.}~\bibnamefont
  {Schramm}},\ }\href {\doibase 10.1103/PhysRevC.66.064310} {\bibfield
  {journal} {\bibinfo  {journal} {Phys. Rev. C}\ }\textbf {\bibinfo {volume}
  {66}},\ \bibinfo {pages} {064310} (\bibinfo {year} {2002})},\ \Eprint
  {http://arxiv.org/abs/nucl-th/0207060} {arXiv:nucl-th/0207060} \BibitemShut
  {NoStop}%
\bibitem [{\citenamefont {Tsubakihara}\ and\ \citenamefont
  {Ohnishi}(2007)}]{Tsubakihara:2006se}%
  \BibitemOpen
  \bibfield  {author} {\bibinfo {author} {\bibfnamefont {K.}~\bibnamefont
  {Tsubakihara}}\ and\ \bibinfo {author} {\bibfnamefont {A.}~\bibnamefont
  {Ohnishi}},\ }\href {\doibase 10.1143/PTP.117.903} {\bibfield  {journal}
  {\bibinfo  {journal} {Prog. Theor. Phys.}\ }\textbf {\bibinfo {volume}
  {117}},\ \bibinfo {pages} {903} (\bibinfo {year} {2007})},\ \Eprint
  {http://arxiv.org/abs/nucl-th/0607046} {arXiv:nucl-th/0607046} \BibitemShut
  {NoStop}%
\bibitem [{\citenamefont {Tsubakihara}\ \emph {et~al.}(2010)\citenamefont
  {Tsubakihara}, \citenamefont {Maekawa}, \citenamefont {Matsumiya},\ and\
  \citenamefont {Ohnishi}}]{Tsubakihara:2009zb}%
  \BibitemOpen
  \bibfield  {author} {\bibinfo {author} {\bibfnamefont {K.}~\bibnamefont
  {Tsubakihara}}, \bibinfo {author} {\bibfnamefont {H.}~\bibnamefont
  {Maekawa}}, \bibinfo {author} {\bibfnamefont {H.}~\bibnamefont {Matsumiya}},
  \ and\ \bibinfo {author} {\bibfnamefont {A.}~\bibnamefont {Ohnishi}},\ }\href
  {\doibase 10.1103/PhysRevC.81.065206} {\bibfield  {journal} {\bibinfo
  {journal} {Phys. Rev. C}\ }\textbf {\bibinfo {volume} {81}},\ \bibinfo
  {pages} {065206} (\bibinfo {year} {2010})},\ \Eprint
  {http://arxiv.org/abs/0909.5058} {arXiv:0909.5058 [nucl-th]} \BibitemShut
  {NoStop}%
\bibitem [{\citenamefont {Boguta}(1983)}]{Boguta:1983uz}%
  \BibitemOpen
  \bibfield  {author} {\bibinfo {author} {\bibfnamefont {J.}~\bibnamefont
  {Boguta}},\ }\href {\doibase 10.1016/0370-2693(83)90065-5} {\bibfield
  {journal} {\bibinfo  {journal} {Phys. Lett. B}\ }\textbf {\bibinfo {volume}
  {128}},\ \bibinfo {pages} {19} (\bibinfo {year} {1983})}\BibitemShut
  {NoStop}%
\bibitem [{\citenamefont {Sahu}\ \emph {et~al.}(1993)\citenamefont {Sahu},
  \citenamefont {Basu},\ and\ \citenamefont {Datta}}]{Sahu:1993db}%
  \BibitemOpen
  \bibfield  {author} {\bibinfo {author} {\bibfnamefont {P.~K.}\ \bibnamefont
  {Sahu}}, \bibinfo {author} {\bibfnamefont {R.}~\bibnamefont {Basu}}, \ and\
  \bibinfo {author} {\bibfnamefont {B.}~\bibnamefont {Datta}},\ }\href
  {\doibase 10.1086/173233} {\bibfield  {journal} {\bibinfo  {journal}
  {Astrophys. J.}\ }\textbf {\bibinfo {volume} {416}},\ \bibinfo {pages} {267}
  (\bibinfo {year} {1993})}\BibitemShut {NoStop}%
\bibitem [{\citenamefont {Jha}\ and\ \citenamefont {Mishra}(2008)}]{Jha2008}%
  \BibitemOpen
  \bibfield  {author} {\bibinfo {author} {\bibfnamefont {T.~K.}\ \bibnamefont
  {Jha}}\ and\ \bibinfo {author} {\bibfnamefont {H.}~\bibnamefont {Mishra}},\
  }\href {\doibase 10.1103/PhysRevC.78.065802} {\bibfield  {journal} {\bibinfo
  {journal} {Phys. Rev.}\ }\textbf {\bibinfo {volume} {C78}},\ \bibinfo {pages}
  {065802} (\bibinfo {year} {2008})},\ \Eprint {http://arxiv.org/abs/0811.4233}
  {arXiv:0811.4233 [nucl-th]} \BibitemShut {NoStop}%
%%CITATION = ARXIV:0811.4233;%%
\bibitem [{\citenamefont {Malik}\ \emph {et~al.}(2017)\citenamefont {Malik},
  \citenamefont {Banerjee}, \citenamefont {Jha},\ and\ \citenamefont
  {Agrawal}}]{Malik2017}%
  \BibitemOpen
  \bibfield  {author} {\bibinfo {author} {\bibfnamefont {T.}~\bibnamefont
  {Malik}}, \bibinfo {author} {\bibfnamefont {K.}~\bibnamefont {Banerjee}},
  \bibinfo {author} {\bibfnamefont {T.~K.}\ \bibnamefont {Jha}}, \ and\
  \bibinfo {author} {\bibfnamefont {B.~K.}\ \bibnamefont {Agrawal}},\ }\href
  {\doibase 10.1103/PhysRevC.96.035803} {\bibfield  {journal} {\bibinfo
  {journal} {Phys. Rev.}\ }\textbf {\bibinfo {volume} {C96}},\ \bibinfo {pages}
  {035803} (\bibinfo {year} {2017})},\ \Eprint
  {http://arxiv.org/abs/1708.07291} {arXiv:1708.07291 [nucl-th]} \BibitemShut
  {NoStop}%
%%CITATION = ARXIV:1708.07291;%%
\bibitem [{\citenamefont {Patra}\ \emph {et~al.}(2020)\citenamefont {Patra},
  \citenamefont {Malik}, \citenamefont {Sen}, \citenamefont {Jha},\ and\
  \citenamefont {Mishra}}]{Patra2020}%
  \BibitemOpen
  \bibfield  {author} {\bibinfo {author} {\bibfnamefont {N.~K.}\ \bibnamefont
  {Patra}}, \bibinfo {author} {\bibfnamefont {T.}~\bibnamefont {Malik}},
  \bibinfo {author} {\bibfnamefont {D.}~\bibnamefont {Sen}}, \bibinfo {author}
  {\bibfnamefont {T.~K.}\ \bibnamefont {Jha}}, \ and\ \bibinfo {author}
  {\bibfnamefont {H.}~\bibnamefont {Mishra}},\ }\href {\doibase
  10.3847/1538-4357/aba8fc} {\bibfield  {journal} {\bibinfo  {journal}
  {Astrophys. J.}\ }\textbf {\bibinfo {volume} {900}},\ \bibinfo {pages} {49}
  (\bibinfo {year} {2020})}\BibitemShut {NoStop}%
\bibitem [{\citenamefont {Bednarek}\ \emph {et~al.}(2012)\citenamefont
  {Bednarek}, \citenamefont {Haensel}, \citenamefont {Zdunik}, \citenamefont
  {Bejger},\ and\ \citenamefont {Manka}}]{Bednarek2012}%
  \BibitemOpen
  \bibfield  {author} {\bibinfo {author} {\bibfnamefont {I.}~\bibnamefont
  {Bednarek}}, \bibinfo {author} {\bibfnamefont {P.}~\bibnamefont {Haensel}},
  \bibinfo {author} {\bibfnamefont {J.}~\bibnamefont {Zdunik}}, \bibinfo
  {author} {\bibfnamefont {M.}~\bibnamefont {Bejger}}, \ and\ \bibinfo {author}
  {\bibfnamefont {R.}~\bibnamefont {Manka}},\ }\href {\doibase
  10.1051/0004-6361/201118560} {\bibfield  {journal} {\bibinfo  {journal}
  {Astron. Astrophys.}\ }\textbf {\bibinfo {volume} {543}},\ \bibinfo {pages}
  {A157} (\bibinfo {year} {2012})},\ \Eprint {http://arxiv.org/abs/1111.6942}
  {arXiv:1111.6942 [astro-ph.SR]} \BibitemShut {NoStop}%
\bibitem [{\citenamefont {Oertel}\ \emph {et~al.}(2015)\citenamefont {Oertel},
  \citenamefont {Providência}, \citenamefont {Gulminelli},\ and\ \citenamefont
  {Raduta}}]{Oertel2015}%
  \BibitemOpen
  \bibfield  {author} {\bibinfo {author} {\bibfnamefont {M.}~\bibnamefont
  {Oertel}}, \bibinfo {author} {\bibfnamefont {C.}~\bibnamefont
  {Providência}}, \bibinfo {author} {\bibfnamefont {F.}~\bibnamefont
  {Gulminelli}}, \ and\ \bibinfo {author} {\bibfnamefont {A.~R.}\ \bibnamefont
  {Raduta}},\ }\href {\doibase 10.1088/0954-3899/42/7/075202} {\bibfield
  {journal} {\bibinfo  {journal} {J. Phys. G}\ }\textbf {\bibinfo {volume}
  {42}},\ \bibinfo {pages} {075202} (\bibinfo {year} {2015})},\ \Eprint
  {http://arxiv.org/abs/1412.4545} {arXiv:1412.4545 [nucl-th]} \BibitemShut
  {NoStop}%
\bibitem [{\citenamefont {Vidana}\ \emph {et~al.}(2000)\citenamefont {Vidana},
  \citenamefont {Polls}, \citenamefont {Ramos}, \citenamefont {Hjorth-Jensen},\
  and\ \citenamefont {Stoks}}]{Vidana2000}%
  \BibitemOpen
  \bibfield  {author} {\bibinfo {author} {\bibfnamefont {I.}~\bibnamefont
  {Vidana}}, \bibinfo {author} {\bibfnamefont {A.}~\bibnamefont {Polls}},
  \bibinfo {author} {\bibfnamefont {A.}~\bibnamefont {Ramos}}, \bibinfo
  {author} {\bibfnamefont {M.}~\bibnamefont {Hjorth-Jensen}}, \ and\ \bibinfo
  {author} {\bibfnamefont {V.~G.~J.}\ \bibnamefont {Stoks}},\ }\href {\doibase
  10.1103/PhysRevC.61.025802} {\bibfield  {journal} {\bibinfo  {journal} {Phys.
  Rev.}\ }\textbf {\bibinfo {volume} {C61}},\ \bibinfo {pages} {025802}
  (\bibinfo {year} {2000})},\ \Eprint {http://arxiv.org/abs/nucl-th/9909019}
  {arXiv:nucl-th/9909019 [nucl-th]} \BibitemShut {NoStop}%
%%CITATION = NUCL-TH/9909019;%%
\bibitem [{\citenamefont {Yamamoto}\ \emph {et~al.}(2014)\citenamefont
  {Yamamoto}, \citenamefont {Furumoto}, \citenamefont {Yasutake},\ and\
  \citenamefont {Rijken}}]{Yamamoto2014}%
  \BibitemOpen
  \bibfield  {author} {\bibinfo {author} {\bibfnamefont {Y.}~\bibnamefont
  {Yamamoto}}, \bibinfo {author} {\bibfnamefont {T.}~\bibnamefont {Furumoto}},
  \bibinfo {author} {\bibfnamefont {N.}~\bibnamefont {Yasutake}}, \ and\
  \bibinfo {author} {\bibfnamefont {T.~A.}\ \bibnamefont {Rijken}},\ }\href
  {\doibase 10.1103/PhysRevC.90.045805} {\bibfield  {journal} {\bibinfo
  {journal} {Phys. Rev.}\ }\textbf {\bibinfo {volume} {C90}},\ \bibinfo {pages}
  {045805} (\bibinfo {year} {2014})},\ \Eprint {http://arxiv.org/abs/1406.4332}
  {arXiv:1406.4332 [nucl-th]} \BibitemShut {NoStop}%
%%CITATION = ARXIV:1406.4332;%%
\bibitem [{\citenamefont {Lonardoni}\ \emph {et~al.}(2013)\citenamefont
  {Lonardoni}, \citenamefont {Gandolfi},\ and\ \citenamefont
  {Pederiva}}]{Lonardoni2013}%
  \BibitemOpen
  \bibfield  {author} {\bibinfo {author} {\bibfnamefont {D.}~\bibnamefont
  {Lonardoni}}, \bibinfo {author} {\bibfnamefont {S.}~\bibnamefont {Gandolfi}},
  \ and\ \bibinfo {author} {\bibfnamefont {F.}~\bibnamefont {Pederiva}},\
  }\href {\doibase 10.1103/PhysRevC.87.041303} {\bibfield  {journal} {\bibinfo
  {journal} {Phys. Rev.}\ }\textbf {\bibinfo {volume} {C87}},\ \bibinfo {pages}
  {041303} (\bibinfo {year} {2013})},\ \Eprint {http://arxiv.org/abs/1301.7472}
  {arXiv:1301.7472 [nucl-th]} \BibitemShut {NoStop}%
%%CITATION = ARXIV:1301.7472;%%
\bibitem [{\citenamefont {Wei}\ \emph {et~al.}(2019)\citenamefont {Wei},
  \citenamefont {Irving}, \citenamefont {Salinas}, \citenamefont {Klähn},\
  and\ \citenamefont {Jaikumar}}]{wei2019}%
  \BibitemOpen
  \bibfield  {author} {\bibinfo {author} {\bibfnamefont {W.}~\bibnamefont
  {Wei}}, \bibinfo {author} {\bibfnamefont {B.}~\bibnamefont {Irving}},
  \bibinfo {author} {\bibfnamefont {M.}~\bibnamefont {Salinas}}, \bibinfo
  {author} {\bibfnamefont {T.}~\bibnamefont {Klähn}}, \ and\ \bibinfo {author}
  {\bibfnamefont {P.}~\bibnamefont {Jaikumar}},\ }\href {\doibase
  10.3847/1538-4357/ab53ea} {\bibfield  {journal} {\bibinfo  {journal} {The
  Astrophysical Journal}\ }\textbf {\bibinfo {volume} {887}},\ \bibinfo {pages}
  {151} (\bibinfo {year} {2019})}\BibitemShut {NoStop}%
\bibitem [{\citenamefont {Kl\"ahn}\ \emph {et~al.}(2013)\citenamefont
  {Kl\"ahn}, \citenamefont {\L{}astowiecki},\ and\ \citenamefont
  {Blaschke}}]{Klahn2013}%
  \BibitemOpen
  \bibfield  {author} {\bibinfo {author} {\bibfnamefont {T.}~\bibnamefont
  {Kl\"ahn}}, \bibinfo {author} {\bibfnamefont {R.}~\bibnamefont
  {\L{}astowiecki}}, \ and\ \bibinfo {author} {\bibfnamefont {D.}~\bibnamefont
  {Blaschke}},\ }\href {\doibase 10.1103/PhysRevD.88.085001} {\bibfield
  {journal} {\bibinfo  {journal} {Phys. Rev. D}\ }\textbf {\bibinfo {volume}
  {88}},\ \bibinfo {pages} {085001} (\bibinfo {year} {2013})}\BibitemShut
  {NoStop}%
\bibitem [{\citenamefont {Rather}\ \emph {et~al.}(2023)\citenamefont {Rather},
  \citenamefont {Marquez}, \citenamefont {Panotopoulos},\ and\ \citenamefont
  {Lopes}}]{Rather:2023dom}%
  \BibitemOpen
  \bibfield  {author} {\bibinfo {author} {\bibfnamefont {I.~A.}\ \bibnamefont
  {Rather}}, \bibinfo {author} {\bibfnamefont {K.~D.}\ \bibnamefont {Marquez}},
  \bibinfo {author} {\bibfnamefont {G.}~\bibnamefont {Panotopoulos}}, \ and\
  \bibinfo {author} {\bibfnamefont {I.}~\bibnamefont {Lopes}},\ }\href
  {\doibase 10.1103/PhysRevD.107.123022} {\bibfield  {journal} {\bibinfo
  {journal} {Phys. Rev. D}\ }\textbf {\bibinfo {volume} {107}},\ \bibinfo
  {pages} {123022} (\bibinfo {year} {2023})},\ \Eprint
  {http://arxiv.org/abs/2303.11006} {arXiv:2303.11006 [nucl-th]} \BibitemShut
  {NoStop}%
\bibitem [{\citenamefont {Rather}\ \emph {et~al.}(2024)\citenamefont {Rather},
  \citenamefont {Marquez}, \citenamefont {Backes}, \citenamefont
  {Panotopoulos},\ and\ \citenamefont {Lopes}}]{Rather:2024hmo}%
  \BibitemOpen
  \bibfield  {author} {\bibinfo {author} {\bibfnamefont {I.~A.}\ \bibnamefont
  {Rather}}, \bibinfo {author} {\bibfnamefont {K.~D.}\ \bibnamefont {Marquez}},
  \bibinfo {author} {\bibfnamefont {B.~C.}\ \bibnamefont {Backes}}, \bibinfo
  {author} {\bibfnamefont {G.}~\bibnamefont {Panotopoulos}}, \ and\ \bibinfo
  {author} {\bibfnamefont {I.}~\bibnamefont {Lopes}},\ }\href {\doibase
  10.1088/1475-7516/2024/05/130} {\bibfield  {journal} {\bibinfo  {journal}
  {JCAP}\ }\textbf {\bibinfo {volume} {05}},\ \bibinfo {pages} {130} (\bibinfo
  {year} {2024})},\ \Eprint {http://arxiv.org/abs/2401.07789} {arXiv:2401.07789
  [nucl-th]} \BibitemShut {NoStop}%
\bibitem [{\citenamefont {Kalita}\ \emph {et~al.}(2024)\citenamefont {Kalita},
  \citenamefont {Routaray}, \citenamefont {Ghosh}, \citenamefont {Kumar},\ and\
  \citenamefont {Agrawal}}]{Kalita:2023rbz}%
  \BibitemOpen
  \bibfield  {author} {\bibinfo {author} {\bibfnamefont {P.~J.}\ \bibnamefont
  {Kalita}}, \bibinfo {author} {\bibfnamefont {P.}~\bibnamefont {Routaray}},
  \bibinfo {author} {\bibfnamefont {S.}~\bibnamefont {Ghosh}}, \bibinfo
  {author} {\bibfnamefont {B.}~\bibnamefont {Kumar}}, \ and\ \bibinfo {author}
  {\bibfnamefont {B.~K.}\ \bibnamefont {Agrawal}},\ }\href {\doibase
  10.1088/1475-7516/2024/04/065} {\bibfield  {journal} {\bibinfo  {journal}
  {JCAP}\ }\textbf {\bibinfo {volume} {04}},\ \bibinfo {pages} {065} (\bibinfo
  {year} {2024})},\ \Eprint {http://arxiv.org/abs/2308.09008} {arXiv:2308.09008
  [nucl-th]} \BibitemShut {NoStop}%
\bibitem [{\citenamefont {Jha}\ \emph {et~al.}(2008)\citenamefont {Jha},
  \citenamefont {Mishra},\ and\ \citenamefont {Sreekanth}}]{Jha2008a}%
  \BibitemOpen
  \bibfield  {author} {\bibinfo {author} {\bibfnamefont {T.~K.}\ \bibnamefont
  {Jha}}, \bibinfo {author} {\bibfnamefont {H.}~\bibnamefont {Mishra}}, \ and\
  \bibinfo {author} {\bibfnamefont {V.}~\bibnamefont {Sreekanth}},\ }\href
  {\doibase 10.1103/PhysRevC.77.045801} {\bibfield  {journal} {\bibinfo
  {journal} {Phys. Rev.}\ }\textbf {\bibinfo {volume} {C77}},\ \bibinfo {pages}
  {045801} (\bibinfo {year} {2008})},\ \Eprint {http://arxiv.org/abs/0710.5392}
  {arXiv:0710.5392 [nucl-th]} \BibitemShut {NoStop}%
%%CITATION = ARXIV:0710.5392;%%
\bibitem [{\citenamefont {Tolman}(1939)}]{Tolman1939}%
  \BibitemOpen
  \bibfield  {author} {\bibinfo {author} {\bibfnamefont {R.~C.}\ \bibnamefont
  {Tolman}},\ }\href {\doibase 10.1103/PhysRev.55.364} {\bibfield  {journal}
  {\bibinfo  {journal} {Phys. Rev.}\ }\textbf {\bibinfo {volume} {55}},\
  \bibinfo {pages} {364} (\bibinfo {year} {1939})}\BibitemShut {NoStop}%
\bibitem [{\citenamefont {Oppenheimer}\ and\ \citenamefont
  {Volkoff}(1939)}]{Oppen1939}%
  \BibitemOpen
  \bibfield  {author} {\bibinfo {author} {\bibfnamefont {J.~R.}\ \bibnamefont
  {Oppenheimer}}\ and\ \bibinfo {author} {\bibfnamefont {G.~M.}\ \bibnamefont
  {Volkoff}},\ }\href {\doibase 10.1103/PhysRev.55.374} {\bibfield  {journal}
  {\bibinfo  {journal} {Phys. Rev.}\ }\textbf {\bibinfo {volume} {55}},\
  \bibinfo {pages} {374} (\bibinfo {year} {1939})}\BibitemShut {NoStop}%
\bibitem [{\citenamefont {{Chandrasekhar}}\ and\ \citenamefont
  {{Ferrari}}(1991)}]{1991RSPSA.434..449C}%
  \BibitemOpen
  \bibfield  {author} {\bibinfo {author} {\bibfnamefont {S.}~\bibnamefont
  {{Chandrasekhar}}}\ and\ \bibinfo {author} {\bibfnamefont {V.}~\bibnamefont
  {{Ferrari}}},\ }\href {\doibase 10.1098/rspa.1991.0104} {\bibfield  {journal}
  {\bibinfo  {journal} {Proceedings of the Royal Society of London Series A}\
  }\textbf {\bibinfo {volume} {434}},\ \bibinfo {pages} {449} (\bibinfo {year}
  {1991})}\BibitemShut {NoStop}%
\bibitem [{\citenamefont {Hinderer}(2008)}]{Hinderer2008}%
  \BibitemOpen
  \bibfield  {author} {\bibinfo {author} {\bibfnamefont {T.}~\bibnamefont
  {Hinderer}},\ }\href {\doibase 10.1086/533487} {\bibfield  {journal}
  {\bibinfo  {journal} {The Astrophysical Journal}\ }\textbf {\bibinfo {volume}
  {677}},\ \bibinfo {pages} {1216} (\bibinfo {year} {2008})}\BibitemShut
  {NoStop}%
\bibitem [{\citenamefont {Hinderer}\ \emph {et~al.}(2010)\citenamefont
  {Hinderer}, \citenamefont {Lackey}, \citenamefont {Lang},\ and\ \citenamefont
  {Read}}]{Hind2010}%
  \BibitemOpen
  \bibfield  {author} {\bibinfo {author} {\bibfnamefont {T.}~\bibnamefont
  {Hinderer}}, \bibinfo {author} {\bibfnamefont {B.~D.}\ \bibnamefont
  {Lackey}}, \bibinfo {author} {\bibfnamefont {R.~N.}\ \bibnamefont {Lang}}, \
  and\ \bibinfo {author} {\bibfnamefont {J.~S.}\ \bibnamefont {Read}},\ }\href
  {\doibase 10.1103/PhysRevD.81.123016} {\bibfield  {journal} {\bibinfo
  {journal} {Phys. Rev. D}\ }\textbf {\bibinfo {volume} {81}},\ \bibinfo
  {pages} {123016} (\bibinfo {year} {2010})}\BibitemShut {NoStop}%
\bibitem [{\citenamefont {Riley}\ \emph {et~al.}(2021)\citenamefont {Riley}
  \emph {et~al.}}]{Riley:2021pdl}%
  \BibitemOpen
  \bibfield  {author} {\bibinfo {author} {\bibfnamefont {T.~E.}\ \bibnamefont
  {Riley}} \emph {et~al.},\ }\href {\doibase 10.3847/2041-8213/ac0a81}
  {\bibfield  {journal} {\bibinfo  {journal} {Astrophys. J. Lett.}\ }\textbf
  {\bibinfo {volume} {918}},\ \bibinfo {pages} {L27} (\bibinfo {year}
  {2021})},\ \Eprint {http://arxiv.org/abs/2105.06980} {arXiv:2105.06980
  [astro-ph.HE]} \BibitemShut {NoStop}%
\bibitem [{\citenamefont {Riley}\ \emph {et~al.}(2019)\citenamefont {Riley}
  \emph {et~al.}}]{Riley:2019yda}%
  \BibitemOpen
  \bibfield  {author} {\bibinfo {author} {\bibfnamefont {T.~E.}\ \bibnamefont
  {Riley}} \emph {et~al.},\ }\href {\doibase 10.3847/2041-8213/ab481c}
  {\bibfield  {journal} {\bibinfo  {journal} {Astrophys. J. Lett.}\ }\textbf
  {\bibinfo {volume} {887}},\ \bibinfo {pages} {L21} (\bibinfo {year}
  {2019})},\ \Eprint {http://arxiv.org/abs/1912.05702} {arXiv:1912.05702
  [astro-ph.HE]} \BibitemShut {NoStop}%
\bibitem [{\citenamefont {Miller}\ \emph {et~al.}(2019)\citenamefont {Miller}
  \emph {et~al.}}]{Miller:2019cac}%
  \BibitemOpen
  \bibfield  {author} {\bibinfo {author} {\bibfnamefont {M.~C.}\ \bibnamefont
  {Miller}} \emph {et~al.},\ }\href {\doibase 10.3847/2041-8213/ab50c5}
  {\bibfield  {journal} {\bibinfo  {journal} {Astrophys. J. Lett.}\ }\textbf
  {\bibinfo {volume} {887}},\ \bibinfo {pages} {L24} (\bibinfo {year}
  {2019})},\ \Eprint {http://arxiv.org/abs/1912.05705} {arXiv:1912.05705
  [astro-ph.HE]} \BibitemShut {NoStop}%
\bibitem [{\citenamefont {Gomes}\ \emph {et~al.}(2015)\citenamefont {Gomes},
  \citenamefont {Dexheimer}, \citenamefont {Schramm},\ and\ \citenamefont
  {Vasconcellos}}]{Gomes2015}%
  \BibitemOpen
  \bibfield  {author} {\bibinfo {author} {\bibfnamefont {R.~O.}\ \bibnamefont
  {Gomes}}, \bibinfo {author} {\bibfnamefont {V.}~\bibnamefont {Dexheimer}},
  \bibinfo {author} {\bibfnamefont {S.}~\bibnamefont {Schramm}}, \ and\
  \bibinfo {author} {\bibfnamefont {C.~A.~Z.}\ \bibnamefont {Vasconcellos}},\
  }\href {\doibase 10.1088/0004-637x/808/1/8} {\bibfield  {journal} {\bibinfo
  {journal} {The Astrophysical Journal}\ }\textbf {\bibinfo {volume} {808}},\
  \bibinfo {pages} {8} (\bibinfo {year} {2015})}\BibitemShut {NoStop}%
\bibitem [{\citenamefont {Schaffner-Bielich}\ and\ \citenamefont
  {Gal}(2000{\natexlab{a}})}]{Gal2000}%
  \BibitemOpen
  \bibfield  {author} {\bibinfo {author} {\bibfnamefont {J.}~\bibnamefont
  {Schaffner-Bielich}}\ and\ \bibinfo {author} {\bibfnamefont {A.}~\bibnamefont
  {Gal}},\ }\href {\doibase 10.1103/PhysRevC.62.034311} {\bibfield  {journal}
  {\bibinfo  {journal} {Phys. Rev. C}\ }\textbf {\bibinfo {volume} {62}},\
  \bibinfo {pages} {034311} (\bibinfo {year} {2000}{\natexlab{a}})}\BibitemShut
  {NoStop}%
\bibitem [{\citenamefont {Friedman}\ and\ \citenamefont
  {Gal}(2021)}]{FRIEDMAN2021136555}%
  \BibitemOpen
  \bibfield  {author} {\bibinfo {author} {\bibfnamefont {E.}~\bibnamefont
  {Friedman}}\ and\ \bibinfo {author} {\bibfnamefont {A.}~\bibnamefont {Gal}},\
  }\href {\doibase https://doi.org/10.1016/j.physletb.2021.136555} {\bibfield
  {journal} {\bibinfo  {journal} {Physics Letters B}\ }\textbf {\bibinfo
  {volume} {820}},\ \bibinfo {pages} {136555} (\bibinfo {year}
  {2021})}\BibitemShut {NoStop}%
\bibitem [{\citenamefont {Schaffner-Bielich}\ and\ \citenamefont
  {Gal}(2000{\natexlab{b}})}]{Schaffner-Bielich:2000igu}%
  \BibitemOpen
  \bibfield  {author} {\bibinfo {author} {\bibfnamefont {J.}~\bibnamefont
  {Schaffner-Bielich}}\ and\ \bibinfo {author} {\bibfnamefont {A.}~\bibnamefont
  {Gal}},\ }\href {\doibase 10.1103/PhysRevC.62.034311} {\bibfield  {journal}
  {\bibinfo  {journal} {Phys. Rev. C}\ }\textbf {\bibinfo {volume} {62}},\
  \bibinfo {pages} {034311} (\bibinfo {year} {2000}{\natexlab{b}})},\ \Eprint
  {http://arxiv.org/abs/nucl-th/0005060} {arXiv:nucl-th/0005060} \BibitemShut
  {NoStop}%
\bibitem [{\citenamefont {Raduta}(2021)}]{Raduta:2021xiz}%
  \BibitemOpen
  \bibfield  {author} {\bibinfo {author} {\bibfnamefont {A.~R.}\ \bibnamefont
  {Raduta}},\ }\href {\doibase 10.1016/j.physletb.2021.136070} {\bibfield
  {journal} {\bibinfo  {journal} {Phys. Lett. B}\ }\textbf {\bibinfo {volume}
  {814}},\ \bibinfo {pages} {136070} (\bibinfo {year} {2021})},\ \Eprint
  {http://arxiv.org/abs/2101.03718} {arXiv:2101.03718 [nucl-th]} \BibitemShut
  {NoStop}%
\bibitem [{\citenamefont {Bedaque}\ and\ \citenamefont
  {Steiner}(2015)}]{Bedaque:2014sqa}%
  \BibitemOpen
  \bibfield  {author} {\bibinfo {author} {\bibfnamefont {P.}~\bibnamefont
  {Bedaque}}\ and\ \bibinfo {author} {\bibfnamefont {A.~W.}\ \bibnamefont
  {Steiner}},\ }\href {\doibase 10.1103/PhysRevLett.114.031103} {\bibfield
  {journal} {\bibinfo  {journal} {Phys. Rev. Lett.}\ }\textbf {\bibinfo
  {volume} {114}},\ \bibinfo {pages} {031103} (\bibinfo {year} {2015})},\
  \Eprint {http://arxiv.org/abs/1408.5116} {arXiv:1408.5116 [nucl-th]}
  \BibitemShut {NoStop}%
\bibitem [{\citenamefont {Moustakidis}\ \emph {et~al.}(2017)\citenamefont
  {Moustakidis}, \citenamefont {Gaitanos}, \citenamefont {Margaritis},\ and\
  \citenamefont {Lalazissis}}]{Moustakidis:2016sab}%
  \BibitemOpen
  \bibfield  {author} {\bibinfo {author} {\bibfnamefont {C.~C.}\ \bibnamefont
  {Moustakidis}}, \bibinfo {author} {\bibfnamefont {T.}~\bibnamefont
  {Gaitanos}}, \bibinfo {author} {\bibfnamefont {C.}~\bibnamefont
  {Margaritis}}, \ and\ \bibinfo {author} {\bibfnamefont {G.~A.}\ \bibnamefont
  {Lalazissis}},\ }\href {\doibase 10.1103/PhysRevC.95.045801} {\bibfield
  {journal} {\bibinfo  {journal} {Phys. Rev. C}\ }\textbf {\bibinfo {volume}
  {95}},\ \bibinfo {pages} {045801} (\bibinfo {year} {2017})},\ \bibinfo {note}
  {[Erratum: Phys.Rev.C 95, 059904 (2017)]},\ \Eprint
  {http://arxiv.org/abs/1608.00344} {arXiv:1608.00344 [nucl-th]} \BibitemShut
  {NoStop}%
\bibitem [{\citenamefont {Tews}\ \emph {et~al.}(2017)\citenamefont {Tews},
  \citenamefont {Lattimer}, \citenamefont {Ohnishi},\ and\ \citenamefont
  {Kolomeitsev}}]{Tews_2017}%
  \BibitemOpen
  \bibfield  {author} {\bibinfo {author} {\bibfnamefont {I.}~\bibnamefont
  {Tews}}, \bibinfo {author} {\bibfnamefont {J.~M.}\ \bibnamefont {Lattimer}},
  \bibinfo {author} {\bibfnamefont {A.}~\bibnamefont {Ohnishi}}, \ and\
  \bibinfo {author} {\bibfnamefont {E.~E.}\ \bibnamefont {Kolomeitsev}},\
  }\href {\doibase 10.3847/1538-4357/aa8db9} {\bibfield  {journal} {\bibinfo
  {journal} {The Astrophysical Journal}\ }\textbf {\bibinfo {volume} {848}},\
  \bibinfo {pages} {105} (\bibinfo {year} {2017})}\BibitemShut {NoStop}%
\bibitem [{\citenamefont {Sanwal}\ \emph {et~al.}(2002)\citenamefont {Sanwal},
  \citenamefont {Pavlov}, \citenamefont {Zavlin},\ and\ \citenamefont
  {Teter}}]{Sanwal_2002}%
  \BibitemOpen
  \bibfield  {author} {\bibinfo {author} {\bibfnamefont {D.}~\bibnamefont
  {Sanwal}}, \bibinfo {author} {\bibfnamefont {G.~G.}\ \bibnamefont {Pavlov}},
  \bibinfo {author} {\bibfnamefont {V.~E.}\ \bibnamefont {Zavlin}}, \ and\
  \bibinfo {author} {\bibfnamefont {M.~A.}\ \bibnamefont {Teter}},\ }\href
  {\doibase 10.1086/342368} {\bibfield  {journal} {\bibinfo  {journal} {The
  Astrophysical Journal}\ }\textbf {\bibinfo {volume} {574}},\ \bibinfo {pages}
  {L61} (\bibinfo {year} {2002})}\BibitemShut {NoStop}%
\bibitem [{\citenamefont {Hambaryan}\ \emph {et~al.}(2017)\citenamefont
  {Hambaryan}, \citenamefont {Suleimanov}, \citenamefont {Haberl},
  \citenamefont {Schwope}, \citenamefont {Neuh\"auser}, \citenamefont {Hohle},\
  and\ \citenamefont {Werner}}]{Hambaryan:2017wvm}%
  \BibitemOpen
  \bibfield  {author} {\bibinfo {author} {\bibfnamefont {V.}~\bibnamefont
  {Hambaryan}}, \bibinfo {author} {\bibfnamefont {V.}~\bibnamefont
  {Suleimanov}}, \bibinfo {author} {\bibfnamefont {F.}~\bibnamefont {Haberl}},
  \bibinfo {author} {\bibfnamefont {A.~D.}\ \bibnamefont {Schwope}}, \bibinfo
  {author} {\bibfnamefont {R.}~\bibnamefont {Neuh\"auser}}, \bibinfo {author}
  {\bibfnamefont {M.}~\bibnamefont {Hohle}}, \ and\ \bibinfo {author}
  {\bibfnamefont {K.}~\bibnamefont {Werner}},\ }\href {\doibase
  10.1051/0004-6361/201630368} {\bibfield  {journal} {\bibinfo  {journal}
  {Astron. Astrophys.}\ }\textbf {\bibinfo {volume} {601}},\ \bibinfo {pages}
  {A108} (\bibinfo {year} {2017})},\ \Eprint {http://arxiv.org/abs/1702.07635}
  {arXiv:1702.07635 [astro-ph.HE]} \BibitemShut {NoStop}%
\bibitem [{\citenamefont {Abbott}\ \emph
  {et~al.}(2020{\natexlab{b}})\citenamefont {Abbott} \emph
  {et~al.}}]{LIGOScientific:2020zkf}%
  \BibitemOpen
  \bibfield  {author} {\bibinfo {author} {\bibfnamefont {R.}~\bibnamefont
  {Abbott}} \emph {et~al.} (\bibinfo {collaboration} {LIGO Scientific,
  Virgo}),\ }\href {\doibase 10.3847/2041-8213/ab960f} {\bibfield  {journal}
  {\bibinfo  {journal} {Astrophys. J. Lett.}\ }\textbf {\bibinfo {volume}
  {896}},\ \bibinfo {pages} {L44} (\bibinfo {year} {2020}{\natexlab{b}})},\
  \Eprint {http://arxiv.org/abs/2006.12611} {arXiv:2006.12611 [astro-ph.HE]}
  \BibitemShut {NoStop}%
\bibitem [{\citenamefont {Andersson}\ and\ \citenamefont
  {Kokkotas}(1996)}]{PhysRevLett.77.4134}%
  \BibitemOpen
  \bibfield  {author} {\bibinfo {author} {\bibfnamefont {N.}~\bibnamefont
  {Andersson}}\ and\ \bibinfo {author} {\bibfnamefont {K.~D.}\ \bibnamefont
  {Kokkotas}},\ }\href {\doibase 10.1103/PhysRevLett.77.4134} {\bibfield
  {journal} {\bibinfo  {journal} {Phys. Rev. Lett.}\ }\textbf {\bibinfo
  {volume} {77}},\ \bibinfo {pages} {4134} (\bibinfo {year}
  {1996})}\BibitemShut {NoStop}%
\bibitem [{\citenamefont {Andersson}\ and\ \citenamefont
  {Kokkotas}(1998)}]{Andersson:1997rn}%
  \BibitemOpen
  \bibfield  {author} {\bibinfo {author} {\bibfnamefont {N.}~\bibnamefont
  {Andersson}}\ and\ \bibinfo {author} {\bibfnamefont {K.~D.}\ \bibnamefont
  {Kokkotas}},\ }\href {\doibase 10.1046/j.1365-8711.1998.01840.x} {\bibfield
  {journal} {\bibinfo  {journal} {Mon. Not. Roy. Astron. Soc.}\ }\textbf
  {\bibinfo {volume} {299}},\ \bibinfo {pages} {1059} (\bibinfo {year}
  {1998})},\ \Eprint {http://arxiv.org/abs/gr-qc/9711088} {arXiv:gr-qc/9711088}
  \BibitemShut {NoStop}%
\bibitem [{\citenamefont {Benhar}\ \emph {et~al.}(2004)\citenamefont {Benhar},
  \citenamefont {Ferrari},\ and\ \citenamefont {Gualtieri}}]{Benhar:2004xg}%
  \BibitemOpen
  \bibfield  {author} {\bibinfo {author} {\bibfnamefont {O.}~\bibnamefont
  {Benhar}}, \bibinfo {author} {\bibfnamefont {V.}~\bibnamefont {Ferrari}}, \
  and\ \bibinfo {author} {\bibfnamefont {L.}~\bibnamefont {Gualtieri}},\ }\href
  {\doibase 10.1103/PhysRevD.70.124015} {\bibfield  {journal} {\bibinfo
  {journal} {Phys. Rev. D}\ }\textbf {\bibinfo {volume} {70}},\ \bibinfo
  {pages} {124015} (\bibinfo {year} {2004})},\ \Eprint
  {http://arxiv.org/abs/astro-ph/0407529} {arXiv:astro-ph/0407529} \BibitemShut
  {NoStop}%
\end{thebibliography}%

\end{document}